# Dual-support smoothed particle hydrodynamics for elastic mechanics

Zili Dai[3], Huilong Ren[3], Xiaoying Zhuang[3,4,*], Timon Rabczuk[1,2,3,*]

[1]*Division of Computational Mechanics, Ton Duc Thang University, Ho Chi Minh City, Viet Nam*
[2]*Faculty of Civil Engineering, Ton Duc Thang University, Ho Chi Minh City, Viet Nam*
[3]*Institute of Structural Mechanics, Bauhaus-Universit at Weimar, Germany*
[4]*Department of Geotechnical Engineering, College of Civil Engineering,Tongji University, China*

**Abstract**

In the standard SPH method, the interaction between two particles might be not pairwise when the support domain varies, which can result in a reduction of accuracy. To deal with this problem, a modified SPH approach is presented in this paper. First of all, a Lagrangian kernel is introduced to eliminate spurious distortions of the domain of material stability, and the gradient is corrected by a linear transformation so that linear completeness is satisfied. Then, concepts of support and dual-support are defined to deal with the unbalanced interactions between the particles with different support domains. Several benchmark problems in one, two and three dimensions are tested to verify the accuracy of the modified SPH model and highlight its advantages over the standard SPH method through comparisons.

# 1 Introduction

Smoothed particle hydrodynamics (SPH) is a mesh-free technique based on a pure Lagrangian description. Originally developed for astrophysical applications by Lucy [35] and Gingold and Monaghan [19], it has been widely adapted by a range of problems in various disciplines [8, 10, 11, 20, 22, 29, 32, 31, 38, 43, 53, 52]. As a meshless technique, the main advantage of SPH method is to bypass the need of a numerical grid to calculate spatial derivatives. Hence, it avoids the problems associated with mesh tangling and distortion, which usually occur in FEM analyses for large deformation impact and explosive loading events. As a Lagrangian technique, it offers advantages in problems with moving boundaries, large deformations, dynamic fracture and multiple phases.

The first SPH application to problems in the framework of solid mechanics was conducted by Libersky and Petschek [26]. Since then, there has been a growing interest in applying SPH to a wide variety of solid mechanics problems with many promising results [7, 9, 18, 21, 30, 41, 43, 44, 45, 61, 62]. While the favorable features of the SPH method and its applications to solid mechanics have been noted, drawbacks, such as inconsistency [3, 40] , tensile instability [21, 57], and zero energy modes [56], have also been identified. Accordingly, various remedies were proposed for these problems and improved the computing accuracy and stability [6, 15, 27, 42, 48, 58].

In SPH, the field variables of each particle are estimated as sums over all neighboring particles which are located in its support domain. In many problems involving plastic deformation or damage, high stress concentrations occur which require a finer discretization with smaller (or varying) support size. When the sizes of support domain change, the interaction of the particles might be unilateral, thus resulting in an unbalanced internal force. Several approaches have been developed in the existing references to deal with this problem [16, 25, 34, 54, 55]. Those approaches are basically based on variable smoothing length or non-uniformed mass distribution to ensure the consistency; the asymmetry of the particle interaction still remains.

This work aims to eliminate the unbalanced internal force in the standard SPH models resulted from varying support domains. Therefore, the concepts of support and dual-support are introduced. The derivation of governing equations in standard SPH method is then modified, and a new SPH model for elastic solid mechanics is established. Several numerical examples are presented to verify the accuracy of the modified SPH model, and its advantages over the standard SPH model are highlighted.

# 2 Traditional SPH approximation

In the traditional SPH method, the entire system is represented by a finite number of particles that carries individual mass. The field variables of those particles are estimated by summing the contributions from the neighboring particles within a certain area:

$$\langle f(x_i)\rangle = \sum_{j\in H_i} \frac{m_j}{\rho_j} f(x_j) W_{ij} \tag{1}$$

$$\langle \nabla \cdot f(x_i)\rangle = \sum_{j\in H_i} \frac{m_j}{\rho_j} f(x_j) \nabla_i W_{ij} \tag{2}$$

where, the angle brackets $\langle\rangle$ denote a particle approximation, $x_i$ represents the concerning particle, and $x_j$ is a neighboring particle in the support area; $\rho$ and $m$ are the density and mass of particles, $W$ is a kernel function.

In most SPH models [12, 13, 14, 27, 49], the kernel function is directly expressed in terms of an Eulerian kernel:

$$W_{ij} = W(\mathbf{x}_i - \mathbf{x}_j, h) \tag{3}$$

where $\mathbf{x}$ is the spatial (Eulerian) coordinate, and $h$ is a parameter that defines the size of the kernel support in the current configuration, known as the smoothing length or dilation parameter. However, as reported by Belytschko et al. [2], instability usually occurs for solid tension in SPH model with an Eulerian kernel. To eliminate the tensile instability, Rabczuk et al. [42] introduced a Lagrangian kernel, which is expressed in terms of material coordinates:

$$W_{ij} = W(\mathbf{X}_i - \mathbf{X}_j, h_t) \tag{4}$$

where $\mathbf{X}$ is the material (Lagrangian) coordinate. For Lagrangian kernels, the neighbors of influence do not change during the course of the simulation, but the domain of influence in the current configuration changes with time. It was shown that SPH formulations based on a Lagrangian kernel eliminate the tensile instability. In Rabczuk et al. [42], a mixed approach was proposed that allows for extremely large deformations.

In this work, We employ the cubic B-spline function, originally used by Monaghan and Lattanzio [39]:

$$W(R,h) = \alpha_d \times \begin{cases} \frac{2}{3} - R^2 + \frac{1}{2}R^3 & 0 \le R < 1 \\ \frac{1}{6}(2-R)^3 & 1 \le R < 2 \\ 0 & R \ge 2 \end{cases} \tag{5}$$

where $\alpha_d$ is the normalization factor, in one-, two- and three-dimensional space, $\alpha_d = 1/h$, $15/7\pi h^2$ and $3/2\pi h^3$, respectively, and $R = \|\mathbf{x}_i - \mathbf{x}_j\|$.

# 3 Corrected SPH algorithm

One drawback of the standard SPH method is its inability to accurately approximate even a constant function when particles are unevenly spaced or located near boundaries. The conditions for the zeroth- and first-order completeness of the SPH approximation are stated as follows:

$$\sum_{j=1}^{N} V_j W_{ij} = 1 \tag{6}$$

$$\sum_{j=1}^{N} V_j W_{ij} \mathbf{X}_j = \mathbf{X}_i \tag{7}$$

The zeroth- and first-order completeness for a derivative of a function are:

$$\sum_{j=1}^{N} V_j \nabla_j W_{ij} = 0 \tag{8}$$

$$\sum_{j=1}^{N} V_j \nabla_j W_{ij} \mathbf{X}_j = 1 \tag{9}$$

In the simple form as stated above, neither of the completeness conditions are fulfilled by the standard SPH approximation. Therefore, a number of correction techniques have been proposed in the literatures [12, 24, 27]. Some works modify the gradient of the kernel to ensure linear completeness for the derivatives, whereas others modify the kernel function itself. In this work, a correction matrix $\mathbf{M}$ proposed by Bonet and Lok [5] is employed to fulfill first-order completeness for the derivatives of a function:

$$\widetilde{\nabla}_i W_{ij} = \mathbf{M}_i \nabla_i W_{ij} \tag{10}$$

The correction matrix $\mathbf{M}$ is obtained at each particle by enforcing that the following e-quation is satisfied by the corrected kernel gradient. This gives

$$\sum_{j=1}^{N_i} V_j (\mathbf{X}_j - \mathbf{X}_i) \otimes \widetilde{\nabla}_i W_{ij} = (\sum_{j=1}^{N_i} V_j (\mathbf{X}_j - \mathbf{X}_i) \otimes \nabla_i W_{ij}) \mathbf{M}_i^T = I \tag{11}$$

from which $\mathbf{M}$ is evaluated explicitly as

$$\mathbf{M}_i = (\sum_{j=1}^{N_i} V_j \nabla_i W_{ij} \otimes (\mathbf{X}_j - \mathbf{X}_i))^{-1} \tag{12}$$

# 4 Dual-support smoothed particle hydrodynamics

## 4.1 The conception of support and dual-support

Distributing the particles evenly with equal mass and support size results in a significant increase in the computation time. Adaptive particle refinement is a good solution to reduce computational cost and improve numerical results, in which support sizes are made to vary from particle to particle [1, 25, 34]. However the interaction of the particles may be unilateral. As shown in Figure 1, particle $x_j$ falls inside the support domain of particle $x_i$, and contributes to the approximation of the field variables of $x_i$. On the other hand, $x_i$ is not inside the support domain of $x_j$, and will not contribute to the discretized summation of $x_j$. Hence, the interaction of the particles is unilateral, thus resulting in an unbalanced internal force. To avoid this problem, we borrow the concept of *dual horizon* proposed by Ren et al. [51] in Peridynamics (PD). The dual horizon formulation guarantees that the interaction between two particles is always pairwise. Since the support size in SPH is not referred to as horizon, we name our method Dual Support SPH (DS-SPH). Note that there are a few contributions showing similarities between certain meshfree methods and PD [4, 17]. As shown later by numerical experiments, it has the potential to eliminate the unbalanced internal force induced by varying support domain. The concept of support and dual-support in SPH is summarized as follows:

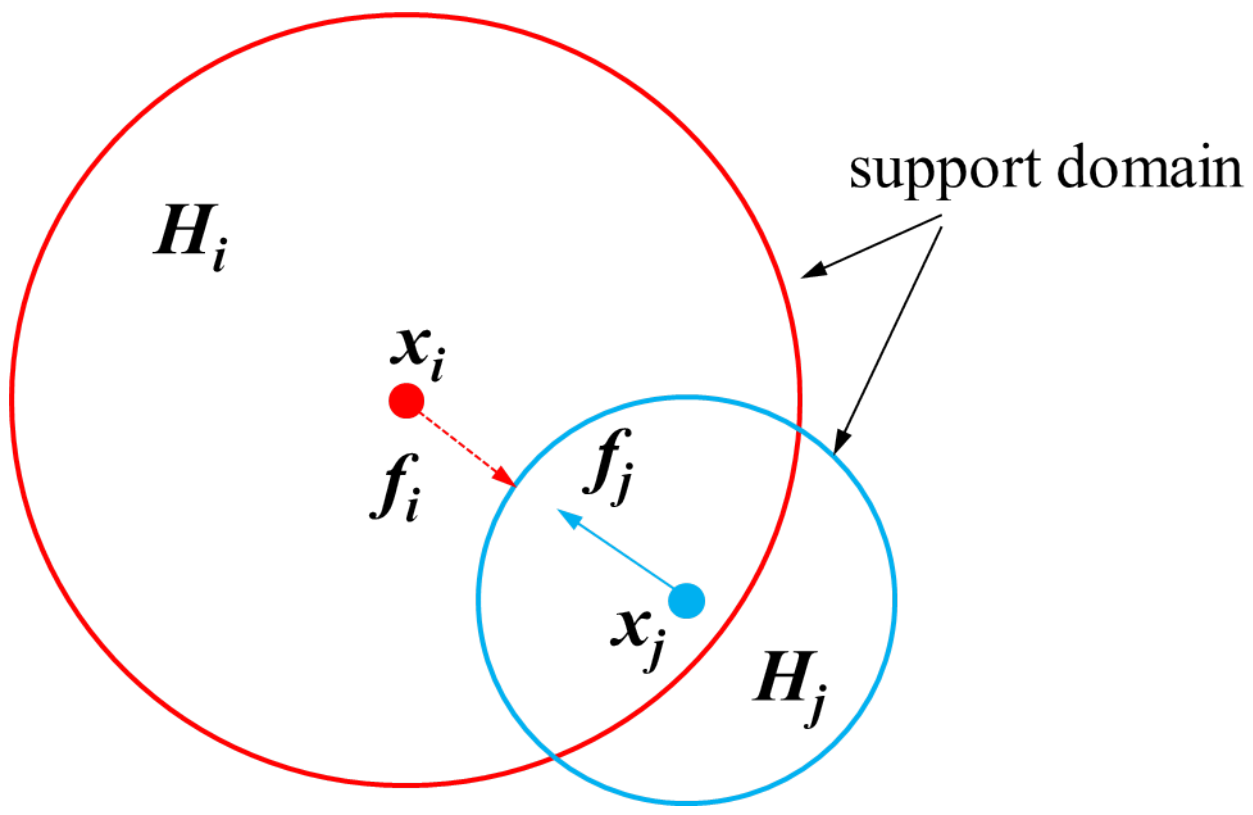

Figure 1: Particle interaction in SPH with varying support domains

The *support* $H_i$ for a field point $x_i$ is the domain where the information for all the points inside this domain is used to determine the information at $x_i$. It is usually centered at particle $i$ with a radius of $h_i$. In Figure 1, $x_j$ is included in the support of $x_i$ because it exerts influence to $x_i$.

The *dual-support* $H'_i$ for a field point $x_i$ is defined as a domain where a particle exerts its influences. Hence, dual-support can be viewed as an *influence domain*. In the notation of dual-support $H'_i$, the superscript prime indicates "dual", and the subscript $i$ denotes the particle that exerts its influences. As shown in Figure 1, $x_j$ exerts influence to $x_i$, so $x_i$ is in its dual-support $H'_j$. On the other hand, $x_i$ makes no contribution to the field variables of $x_j$, so $x_j$ is not included in $H'_i$.

From the definition of support and dual-support discussed above, it is easy to obtain that, the particle $x_i$ is not always included in $H_j$ though particle $x_j$ is in $H_i$, as shown in Figure 1. However, the particle $x_i$ must be included in the dual-support of $x_j$ as long as $x_j$ falls into the support of $x_i$. That means: if $x_j \in H_i$, then $x_i \in H'_j$. Therefore, using the support and dual-support jointly in the SPH model can avoid the unbalanced internal force caused by varying support, because the interaction between any two particles is always pairwise.

## 4.2 Dual property of SPH formulations

Assuming that $F(i,j)$ is the direct force acting on particle $x_i$ due to particle $x_j$. For any domain under consideration $\Omega$, the internal force can be calculated by two approaches. The first approach is by summing all the forces the particles undertake in their support:

$$\sum_{j\in\Omega}\sum_{i\in H_j} F(j,i)\Delta V_i \Delta V_j \tag{13}$$

And the second is to add up all the forces of each particle that it applies to other particles in its dual-support:

$$\sum_{j\in\Omega}\sum_{i\in H'_j} F(i,j)\Delta V_i \Delta V_j \tag{14}$$

Combining equation (13) and (14):

$$\sum_{j\in\Omega}\sum_{i\in H_j} F(j,i)\Delta V_i \Delta V_j = \sum_{j\in\Omega}\sum_{i\in H'_j} F(i,j)\Delta V_i \Delta V_j \tag{15}$$

Therefore, the dual property of the DS-SPH can be stated as that the summation of the term $F(j,i)$ in support can be converted to the summation of the term $F(i,j)$ in dual-support. The particle volumes $\Delta V_i$ and $\Delta V_j$ can be replaced with mass $m_i$ and $m_j$ respectively.
Based on the dual property, we can convert the standard SPH formulation into dual-support formulation. For example, assuming $f(\mathbf{x}_i)$ is a field function on particle $x_i$, $\rho_i$ is the density, in standard SPH formulation,$\nabla f(\mathbf{x}_i)/\rho_i$ is expressed as:

$$\frac{\nabla f(\mathbf{x}_i)}{\rho_i} = \sum_{j\in H_i} \frac{f(\mathbf{x}_j)}{\rho_j^2} m_j \nabla_i W_{ij} - \sum_{j\in H_i} \frac{f(\mathbf{x}_i)}{\rho_i^2} m_j \nabla_i W_{ij} \tag{16}$$

Then in dual-support SPH formulation, the above equation can be converted into:

$$\frac{\nabla f(\mathbf{x}_i)}{\rho_i} = -\sum_{j\in H'_i} \frac{f(\mathbf{x}_j)}{\rho_j^2} m_j \nabla_j W_{ij} - \sum_{j\in H_i} \frac{f(\mathbf{x}_i)}{\rho_i^2} m_j \nabla_i W_{ij} \tag{17}$$

According to this property, the governing equations can be rewritten in dual-support SPH formulation, as shown in Section 5.2.

# 5 Dual-support Smoothed Particle Hydrodynamics (DS-SPH)

## 5.1 Governing equations

The conservation equations for mass, momentum and energy for solid mechanics problems neglecting body forces and temperature in this work are given by

$$\rho J = \rho_0 J_0 \tag{18}$$

$$\ddot{\mathbf{u}} = \frac{1}{\rho_0} \nabla_0 \cdot \mathbf{P} \tag{19}$$

$$\dot{e} = \frac{1}{\rho_0} \mathbf{P} : \dot{\mathbf{F}} \tag{20}$$

where $J$ and $J_0$ are the current and initial Jacobian determinants respectively; $\rho$ and $\rho_0$ are the current and initial mass density, $\mathbf{P}$ is the first Piola-Kirchhoff stress tensor and $e$ is the internal energy density and $\dot{\mathbf{F}}$ denotes the incremental deformation gradient.

The Jacobian determinant can be obtained by

$$J = det(\mathbf{F}) \tag{21}$$

where $\mathbf{F}$ is the deformation gradient which is defined as:

$$\mathbf{F} = \frac{d\mathbf{x}}{d\mathbf{X}} = \frac{d\mathbf{u}}{d\mathbf{X}} + \mathbf{I} \tag{22}$$

With the deformation gradient in hand, the Green-Lagrange strain can be easily obtained by:

$$\mathbf{E} = \frac{1}{2}(\mathbf{F}^T\mathbf{F} - \mathbf{I}) \tag{23}$$

where $\mathbf{F}^T$ is the transposition of deformation gradient.
Linear elasticity constitutive model is selected in this work, and the second Piola-Kirchhoff stress $\mathbf{S}$ is given by

$$\mathbf{S} = \lambda Tr\{\mathbf{E}\} + 2\mu\mathbf{E} \tag{24}$$

where $\lambda$ and $\mu$ are lame constants.
Then the first Piola-Kirchhoff stress $\mathbf{P}$ is given by

$$\mathbf{P} = \mathbf{FS} \tag{25}$$

## 5.2 SPH discrete form of governing equations

According to the corrected SPH algorithm, the approximated deformation gradient at a material point $x_i$ is given by

$$\mathbf{F}_i = \sum_{j\in H_i} \frac{m_j}{\rho_j} (\mathbf{u}_j - \mathbf{u}_i) \otimes \widetilde{\nabla}_i W_{ij} + \mathbf{I} \tag{26}$$

Similarly, the momentum equation and energy equation are given by [28]:

$$\frac{d\mathbf{v}_i}{dt} = \sum_{j\in H_i} m_j(\frac{\mathbf{P}_i}{\rho_i^2} + \frac{\mathbf{P}_j}{\rho_j^2})\widetilde{\nabla}_i W_{ij} \tag{27}$$

$$\frac{de_i}{dt} = \sum_{j\in H_i} m_j(\frac{\mathbf{P}_i}{\rho_i^2} + \frac{\mathbf{P}_j}{\rho_j^2})\mathbf{v}_{ij}\widetilde{\nabla}_i W_{ij} \tag{28}$$

In the above conservation equations, to avoid the unbalanced internal force caused by the varying support, the dual-support conception is introduced. According to the dual property of the DS-SPH discussed in Section 4.2, equation (27) and (28) can be rewritten as:

$$\frac{d\mathbf{v}_i}{dt} = \sum_{j\in H_i} \frac{\mathbf{P}_i}{\rho_i^2}\widetilde{\nabla}_i W_{ij} m_j - \sum_{j\in H_i'} \frac{\mathbf{P}_j}{\rho_j^2}\widetilde{\nabla}_j W_{ij} m_j \tag{29}$$

$$\frac{de_i}{dt} = \sum_{j\in H_i} \frac{\mathbf{P}_i}{\rho_i^2}\mathbf{v}_{ij}\widetilde{\nabla}_i W_{ij} m_j - \sum_{j\in H_i'} \frac{\mathbf{P}_j}{\rho_j^2}\mathbf{v}_{ij}\widetilde{\nabla}_j W_{ij} m_j \tag{30}$$

In order to avoid the unphysical oscillations in the numerical results and then improve the numerical stability, the Monaghan type artificial viscosity [37] is added in the momentum equation. The original form of the artificial viscosity is given by:

$$\Pi_{ij} = \frac{-a\bar{c}_{ij}\mu_{ij} + b(\mu_{ij})^2}{\bar{\rho}_{ij}} \tag{31}$$

where $\bar{\rho}_{ij}$ and $\bar{c}_{ij}$ are the average of the density and sound speed between particle $i$ and $j$. $\mu_{ij}$ can be defined as

$$\mu_{ij} = \frac{h_{ij}\mathbf{v}_{ij}\mathbf{r}_{ij}}{\mathbf{r}_{ij}^2 + 0.01h_{ij}} \tag{32}$$

where $h_{ij}$ is the average smoothing length, $\mathbf{v}_{ij}$ and $\mathbf{r}_{ij}$ are the relative velocity and position respectively.
In the presented dual support SPH model, the artificial viscosity is rewritten as

$$\Pi_{ij} = \frac{-ac_i\mu_{ij} + b(\mu_{ij})^2}{2\rho_i} \tag{33}$$

$$\mu_{ij} = \frac{h_i\mathbf{v}_{ij}\mathbf{r}_{ij}}{\mathbf{r}_{ij}^2 + 0.01h_i} \tag{34}$$

Therefore, the momentum equation and energy equation in DS-SPH model are:

$$\frac{d\mathbf{v}_i}{dt} = \sum_{j\in H_i} (\frac{\mathbf{P}_i}{\rho_i^2} + \delta\Pi_{ij})\widetilde{\nabla}_i W_{ij} m_j - \sum_{j\in H_i} (\frac{\mathbf{P}_j}{\rho_j^2} + \delta\Pi_{ji})\widetilde{\nabla}_j W_{ij} m_j \tag{35}$$

$$\frac{de_i}{dt} = \sum_{j\in H_i} (\frac{\mathbf{P}_i}{\rho_i^2} + \delta\Pi_{ij})\mathbf{v}_{ij}\widetilde{\nabla}_i W_{ij} m_j - \sum_{j\in H_i} (\frac{\mathbf{P}_j}{\rho_j^2} + \delta\Pi_{ji})\mathbf{v}_{ij}\widetilde{\nabla}_j W_{ij} m_j \tag{36}$$

# 6 Numerical examples

In this section, the dual-support SPH method outlined above is applied and tested. The solutions to benchmark problems in one, two and three dimensions are provided, and comparisons with analytical or standard SPH analysis are conducted.

## 6.1 Longitudinal vibration of a bar

The first benchmark is the longitudinal vibration of a bar [36]. A one-dimensional bar is subjected to an initial stretch, and then the stretch is released after a short period of time. As illustrated in Figure 2, the bar is fixed at the left end, and free at the right end. The solution is obtained by specifying the geometric parameters, material properties, as well as the initial and boundary conditions. The total length of the bar is 1m. The mass density, Young's modulus and Poisson's ratio used in the simulation are: $\rho = 7850kg/m^3$, $E = 200GPa$, $\nu = 0.25$. The initial velocity of each material point is $v_x(x,t) = 0$. The initial displacement gradient is defined as following:

$$\frac{\partial u_x}{\partial x} = \varepsilon H(\triangle t - t) \tag{37}$$

where $H$ is a step function; $\varepsilon = 0.001$ is the initial strain.

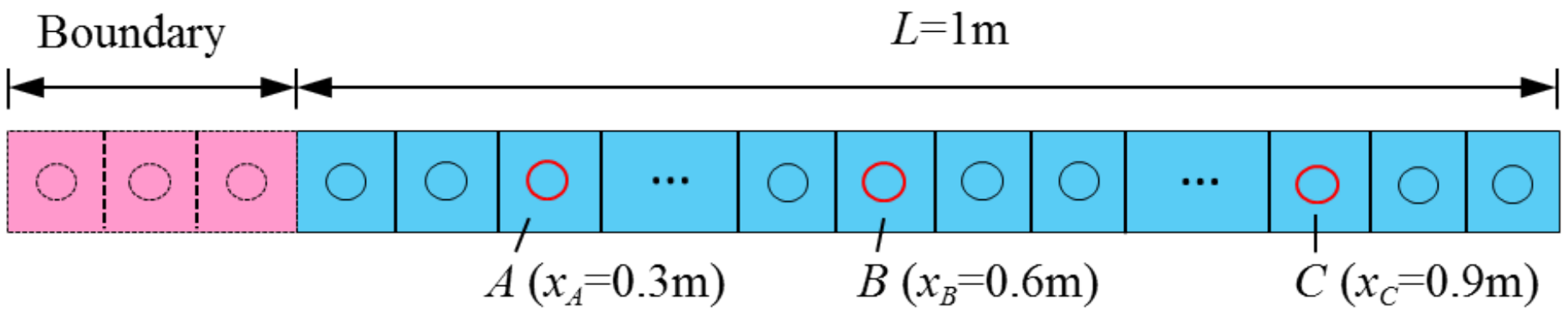

Figure 2: Geometry of a bar subjected to initial strain and its discretization

The proposed DS-SPH model is used to simulate the longitudinal vibration of a bar. The bar is discretized with 1003 particles with initial spacing of 0.001m; including 3 fixed boundary particles at the left side of the bar (see Figure 2). The support radius associated to each particle is set as 3 times the particle size. According to Rao and Yap [50], the analytical solution to this problem can be easily constructed by the following equation:

$$u_x(x,t) = \frac{8\varepsilon L}{\pi^2} \sum_{n=0}^{\infty} \frac{(-1)^n}{(2n+1)^2} \sin[\frac{(2n+1)\pi x}{2}] \cos[\sqrt{\frac{E}{\rho}} \frac{(2n+1)\pi}{2} t] \tag{38}$$

As shown in Figure 2, three material points A, B, C located at $x_A = 0.3m$, $x_B = 0.6m$ and $x_C = 0.9m$ are monitored, and their displacement variations with time are compared against the analytical solutions. As shown in Figure 3, it is evident that the DS-SPH simulation results are consistent with the analytic solution, thus verifying the accuracy of the proposed DS-SPH model.

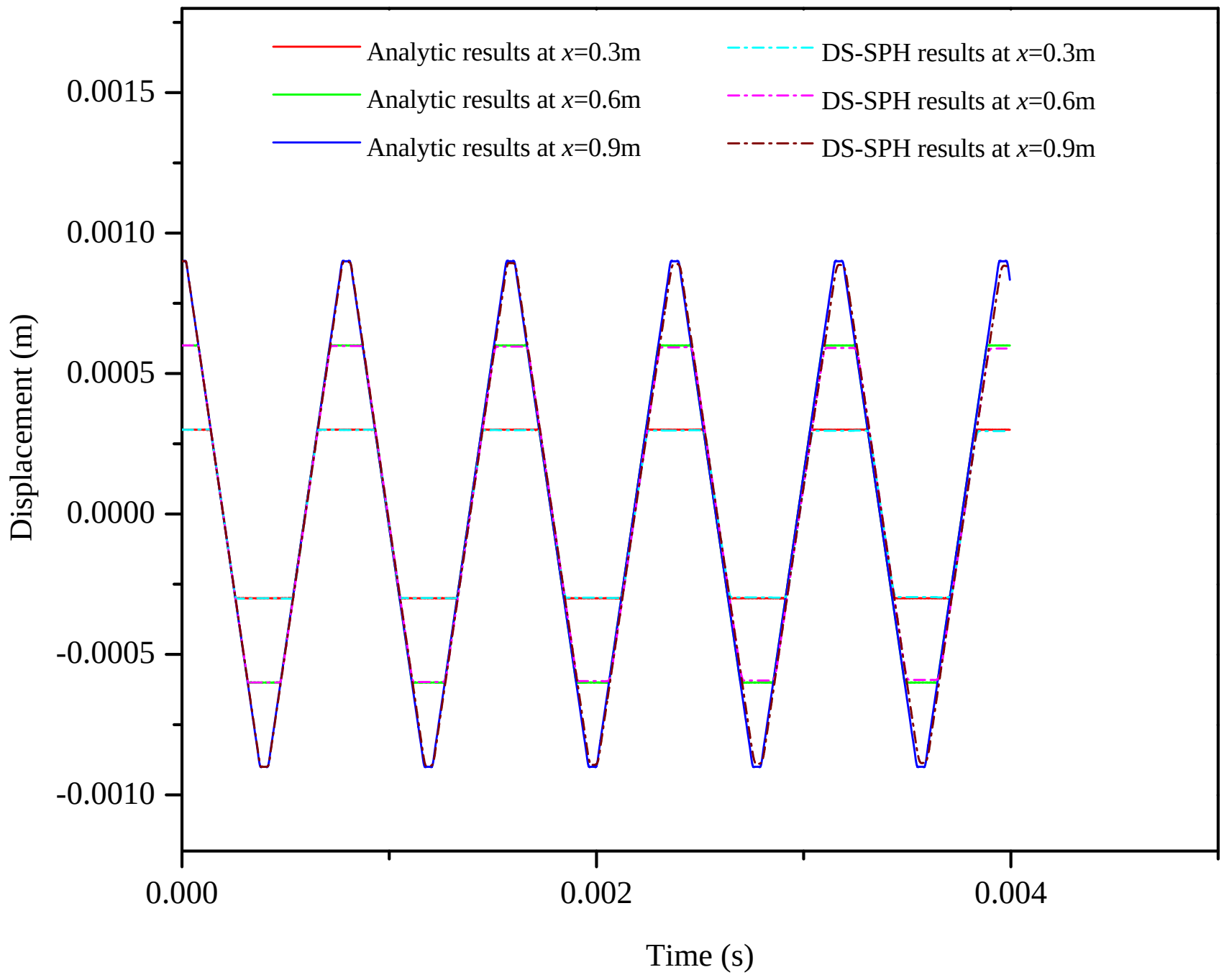

Figure 3: Displacement of a material point located at different positions as a function of time

To test the performance of the DS-SPH model on uneven particle distribution problems, the bar is discretized into 1403 particles with two different particle radiuses. As shown in Figure 4, the bar is divided into one dense zone in the middle of the bar and two coarse zones at the both ends. In the dense zone, a series of fine particles with the radius of 0.0005m are distributed, while the coarse zones are represented by particles with radius of 0.001m. The support radius associated to each particle is set as 3 times the particle size. Therefore, along the interface between the coarse and fine discretization, the support sizes vary.

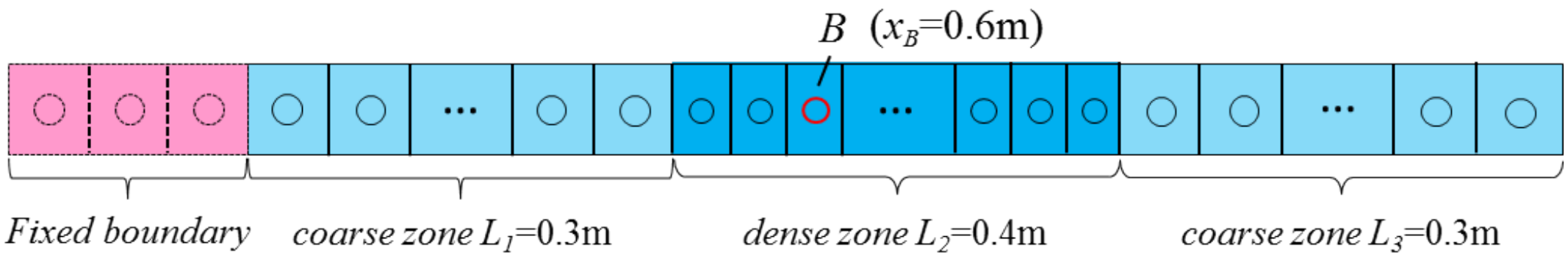

Figure 4: Geometry of a bar subjected to initial strain and its discretization

For comparison purposes, both traditional SPH model and DS-SPH model are used here to simulate the vibration of the bar. The material properties, initial and boundary conditions are exactly the same as the example presented above. The time history of the displacement at point B ($x_B = 0.6m$) are presented in Figure 5. The red solid line is the analytic resolution based on the equation (38); the green and blue dot lines are respectively the numerical results from the SPH and DS-SPH models. It is easy to see that the DS-SPH results (the blue dot line) agree well with the theoretical value, while the SPH results (the green dot line) gradually deviate from the theoretical value. Figure 6 shows the displacement distribution along the bar at $t = 0.325s$. The blue line represents SPH results and the red one is the DS-SPH results. Near the interface between the coarse and fine discretization ($x = 0.3m$ and $x = 0.7m$), the displacement from SPH model is found to fluctuate considerably, while the DS-SPH result is relatively stable.

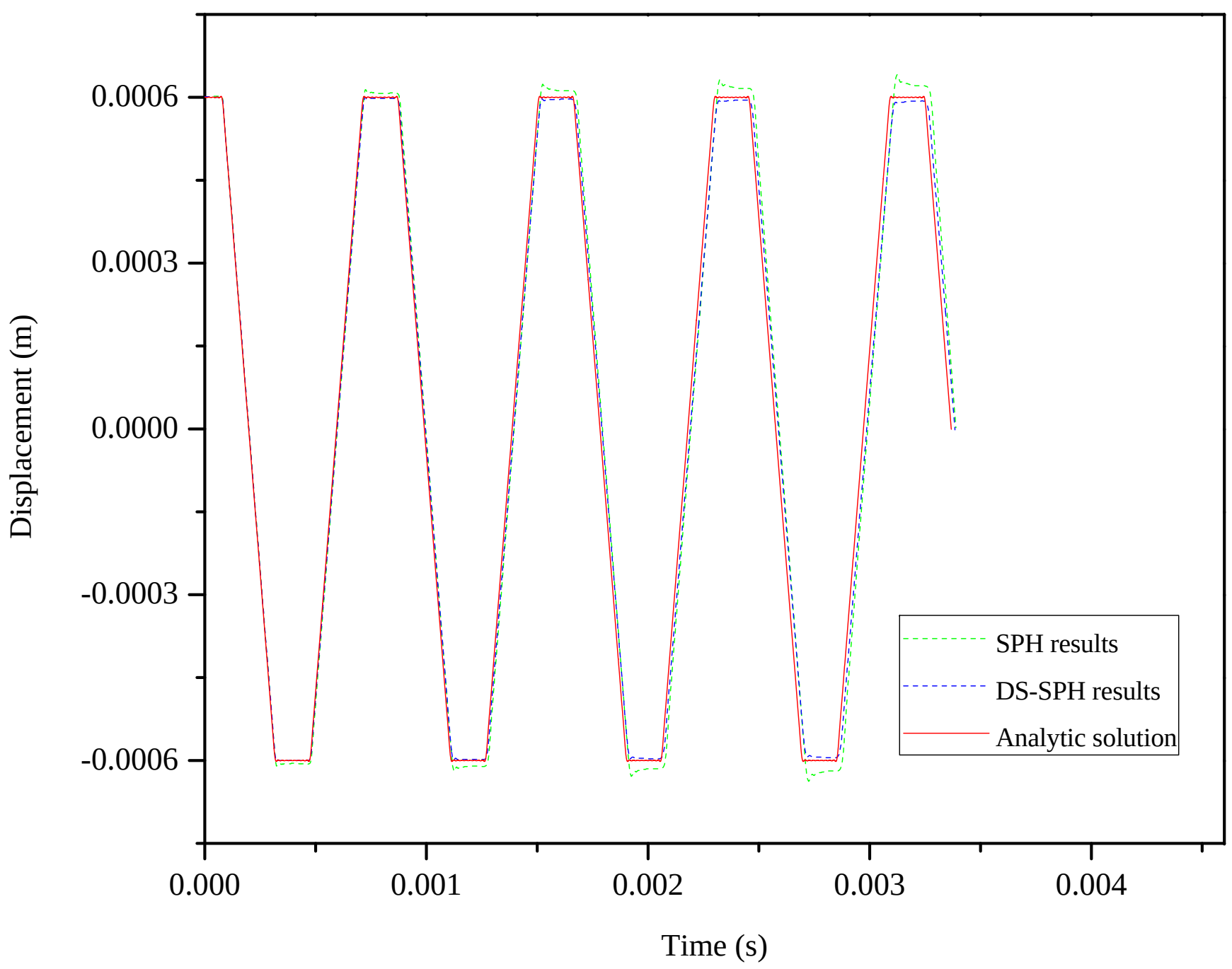

Figure 5: Displacement of a material point located at $x = 0.6m$ as a function of time

To quantitatively compare the accuracy of the standard SPH and proposed DS-SPH model, the $L_2$ error in the displacement field is evaluated, which is given by:

$$\parallel err \parallel_{L_2} = \frac{\parallel \mathbf{u}' - \mathbf{u}_a \parallel}{\parallel \mathbf{u}_a \parallel} \tag{39}$$

where $\mathbf{u}'$ is the numerical displacement and $\mathbf{u}_a$ is the analytical value. The norm $\parallel \mathbf{u} \parallel$ can be calculated as:

$$\parallel \mathbf{u} \parallel_{L_2} = (\int_{\Omega_0} \mathbf{u} \cdot \mathbf{u} d\Omega_0)^{1/2} \tag{40}$$

Table 1 presents the $L_2$ error of displacement to show the numerical accuracy of the two models. For the problem with varying particle support domain, the computational accuracy of the DS-SPH model is better than the standard SPH model.

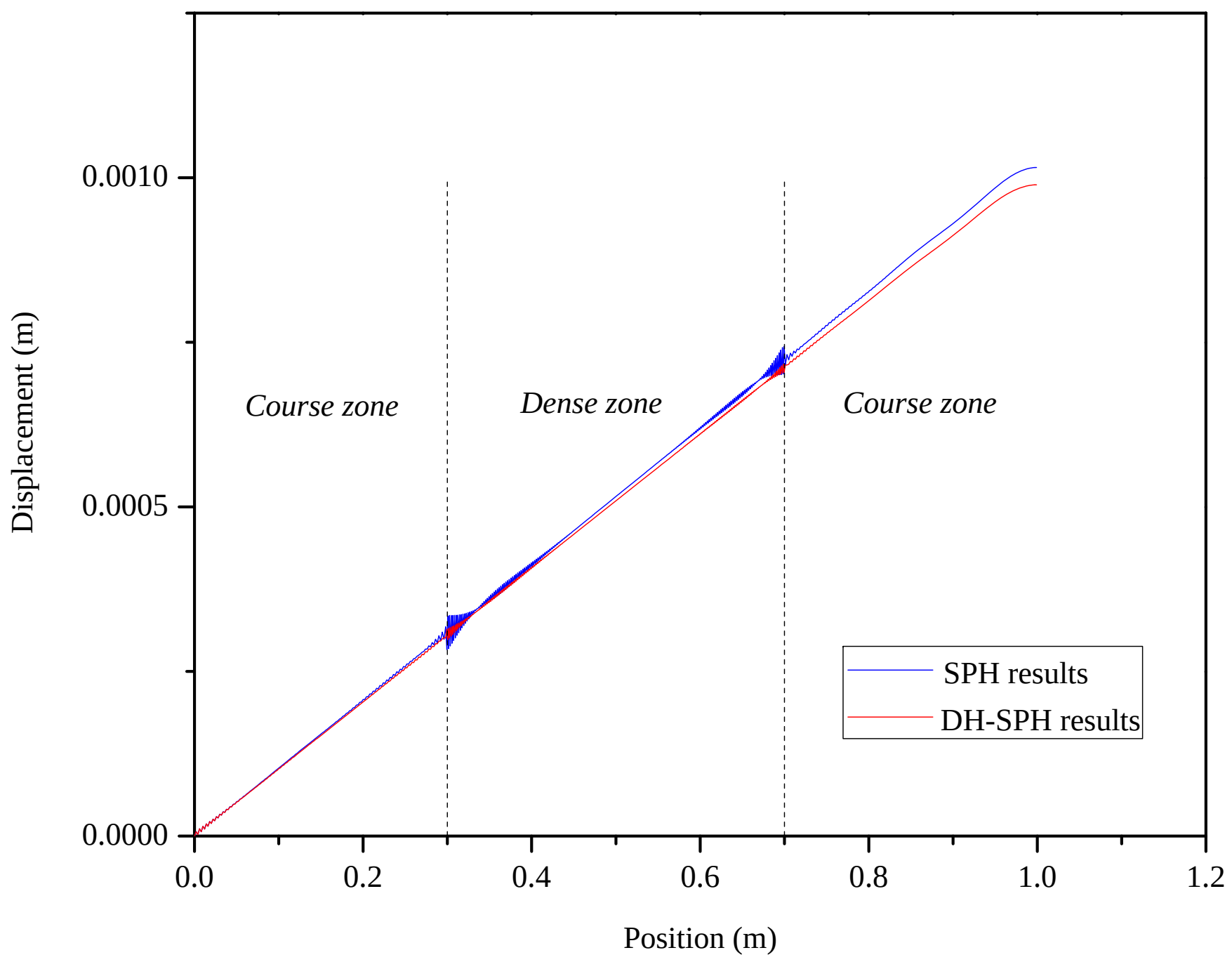

Figure 6: Displacement distribution along the bar at $t = 0.325s$

Table 1: $L_2$ error of displacement at different times

| Model | $t = 0.8ms$ | $t = 1.6ms$ | $t = 2.4ms$ | $t = 3.2ms$ |
|---|---|---|---|---|
| SPH | 0.0104 | 0.0182 | 0.0261 | 0.0340 |
| DS-SPH | 0.0045 | 0.0081 | 0.0118 | 0.0156 |

## 6.2 Wave reflection in a rectangular plate

The second numerical example is the wave reflection in a rectangular plate. Figure 7 shows a rectangular plate with dimensions of $0.1 \times 0.04m^2$. The density, Young's modulus and Poisson's ratio of the material are assumed as $\rho = 1kg/m^3$, $E = 1Pa$ and $\nu = 0$, respectively. Then the wave speed is $v = (E/\rho)^{1/2} = 1m/s$. The initial state of the plate is described by the following equations:

$$u_0(x, y) = 0.0002exp[-(\frac{x}{0.01})^2] \tag{41}$$

$$v_0(x, y) = 0.0 \tag{42}$$

where $u_0$ and $v_0$ denote the displacement and velocity in the $x$ and $y$ directions respectively.

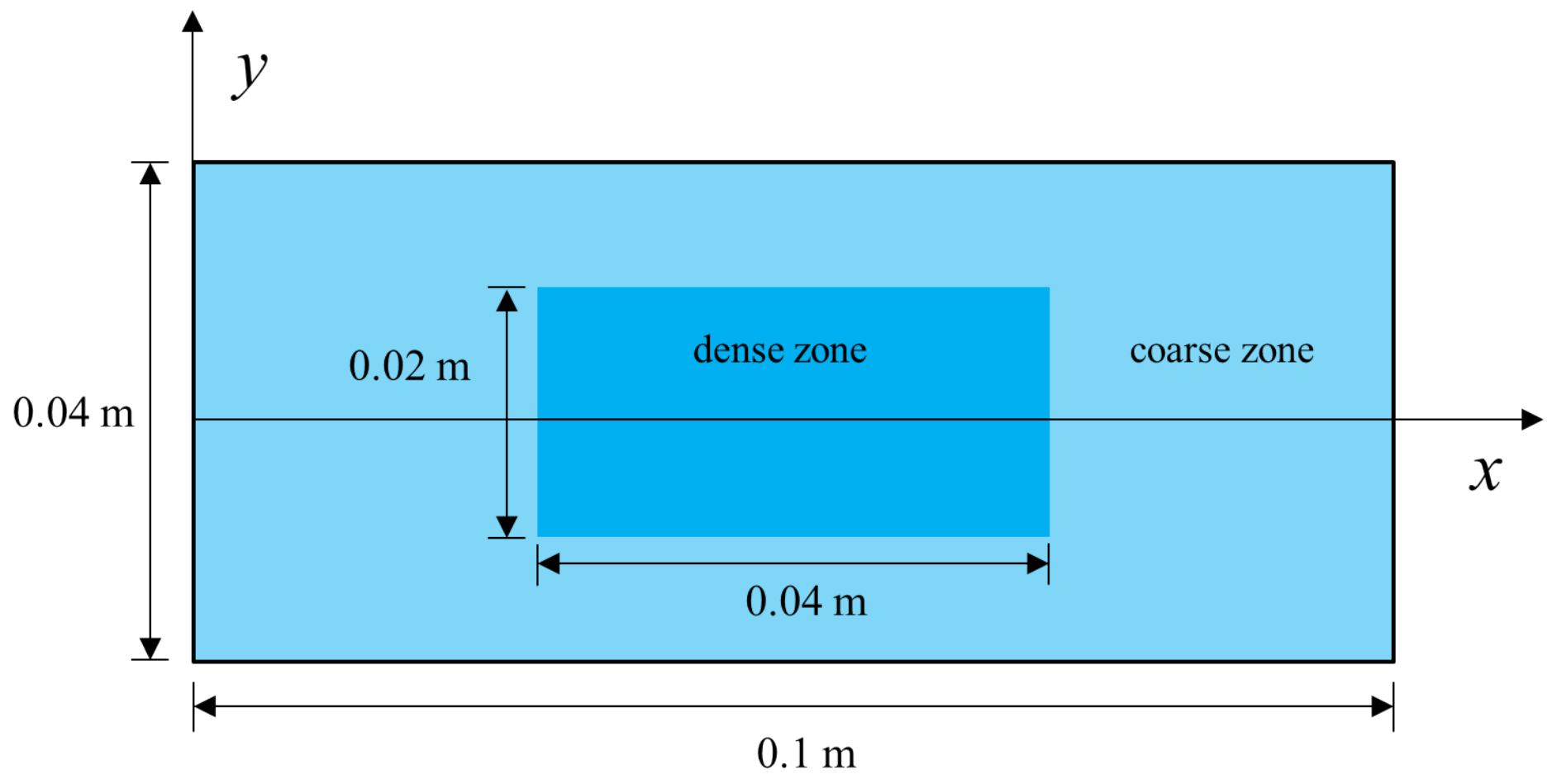

Figure 7: Setup of the plate

As shown in Table 2, three models for solving this problem, namely model A, B and C, are devised. In model A, the plate is discretised with 4000 particles with the same diameter of 0.001 m. For models B and C, we adopted a discretisation of varying particle sides, 3200 fine particles with diameter of 0.0005 m and 3200 coarse particles with diameter of 0.001 m. The support domains associated to each particle are set as 3 times the particle sizes for all models. Therefore, along the interface between the coarse and fine discretization, the support domains vary. Model A and Model B are solved with the standard SPH method with constant and variable support domains respectively. Model C is the dual-support formulation for SPH with variable support domains.

Table 2: Three models for wave reflection in a rectangular plate

| Model | $\Delta x = \Delta y$ | Particle numbers | Numerical approach |
|---|---|---|---|
| A | 0.002 | 4000 | SPH |
| B | 0.001,0.002 | 6400 | SPH |
| C | 0.001,0.002 | 6400 | DS-SPH |

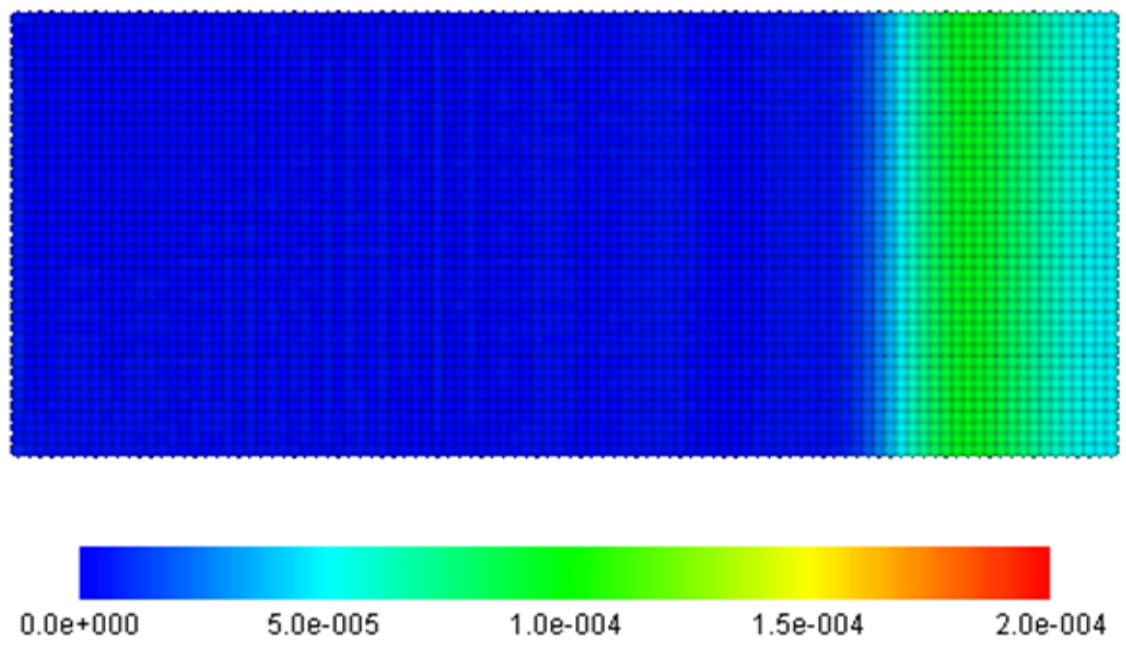

(a) Model A

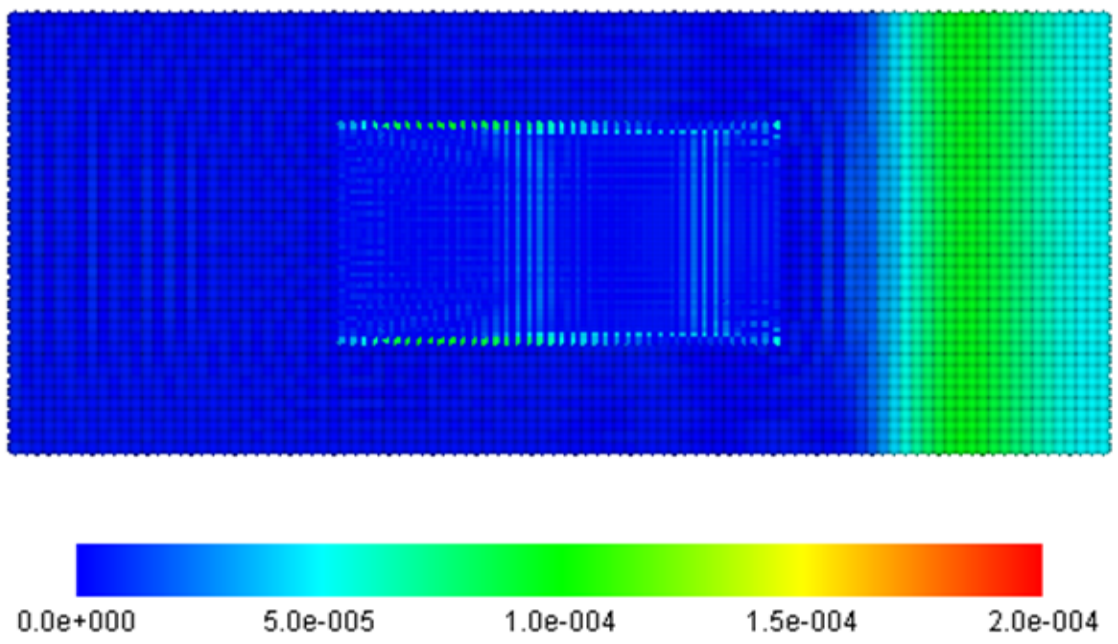

(b) Model B

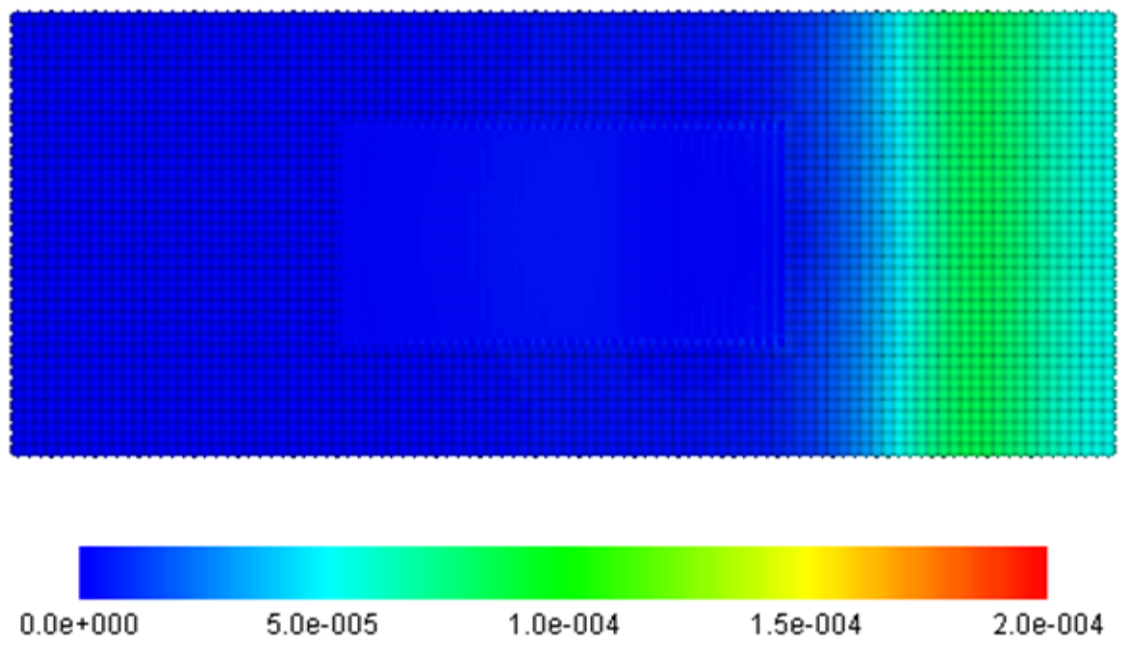

(c) Model C

Figure 8: The displacement contour at $t = 0.09s$

Figure 8 shows the displacement contour in the rectangular plate calculated by the three models mentioned above. The displacement in model C (DS-SPH with variable support domain) is almost identical to that of model A (standard SPH with constant support domain), as shown in Figure 8 (a) and (c). For model B, spurious wave reflections are observed near the interface between the coarse and fine discretization, as shown in Figure 8 (b).

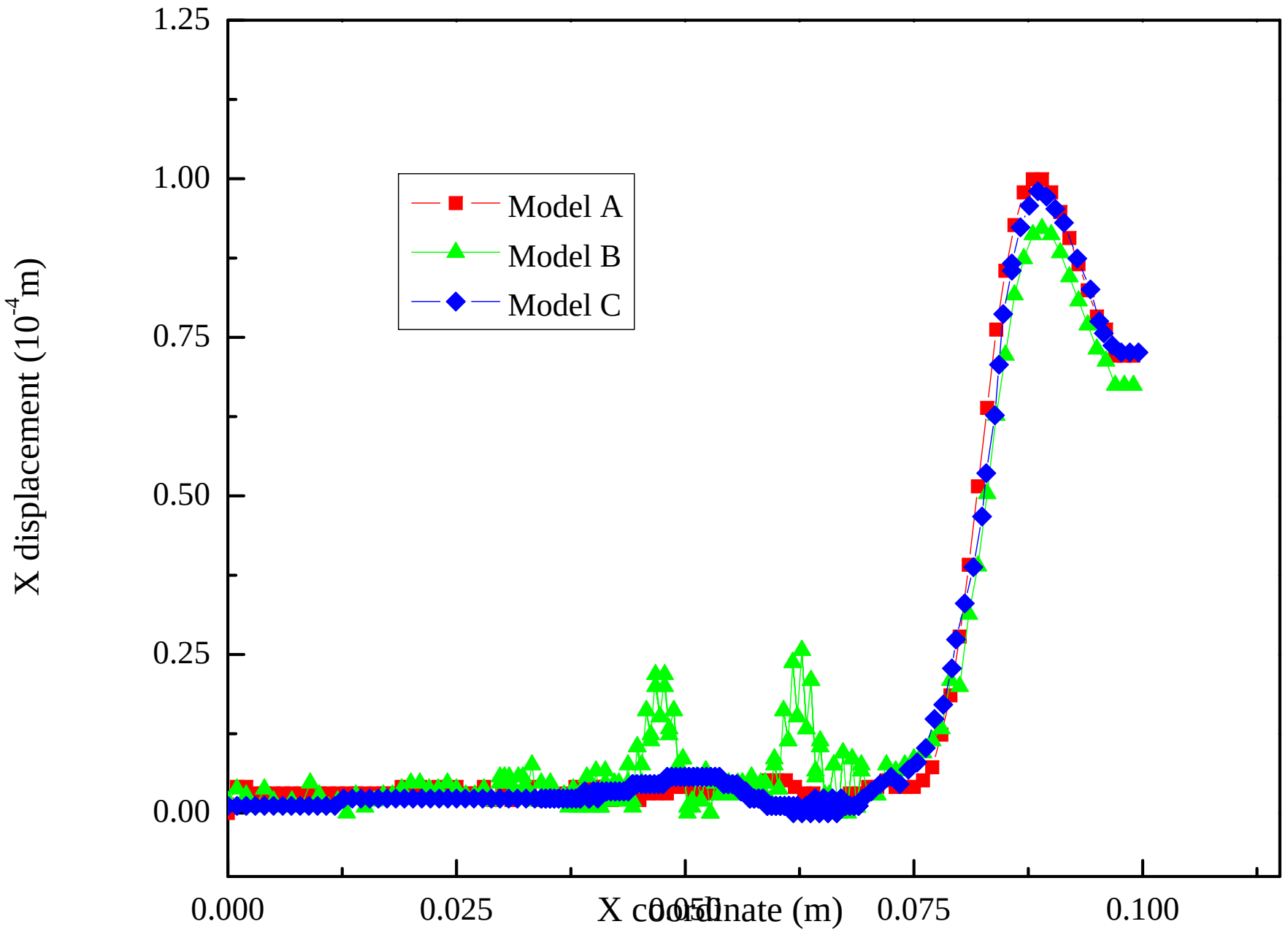

Figure 9: Displacement wave $u_x$ versus $x$ coordinate at $t = 0.09s$

Figure 9 compares the displacements of all particles along the $x$-coordinate calculated by the three models. The results from model A (red squares) and model C (blue diamonds) nearly coincide. The peak value of the displacement is nearly $1.0 \times 10^{-4}m$ which matches the analytical solution well. For model B, two spurious waves are obtained at the position of $x = 0.047m$ and $x = 0.063m$. The peak displacement is $0.92 \times 10^{-4}m$, which is lower than that of Models A and C.

Table 3 presents the $L_2$ error in the displacement to show the numerical accuracy of the three models. The present DS-SPH can achieve almost the same accuracy as the standard SPH model with constant support domain.

Table 3: $L_2$ error of displacement at different times

| Model | $t = 0.1s$ | $t = 0.2s$ | $t = 0.3s$ | $t = 0.4s$ |
|---|---|---|---|---|
| A | 0.0799 | 0.0922 | 0.1204 | 0.1839 |
| B | 0.1968 | 0.2102 | 0.2929 | 0.4419 |
| C | 0.0857 | 0.1261 | 0.1615 | 0.2131 |

## 6.3 Isotropic plate with a circular hole

Let us consider an isotropic plate with a circular hole under uniaxial tension. Stress concentrations occur near the hole, while the stress in the other part of the plate is distributed evenly. Therefore, fine meshes are necessary to analyze the stress near the hole in the numerical simulation. To further test the performance on the uneven particle distribution problems, the DS-SPH model is used to analyze the stress distribution of the isotropic plate with a circular cutout under uniaxial tension.

The geometry of the plate and the cutout are shown in Figure 10. The length and width of the rectangular isotropic plate are $0.10m$; the radius of the cutout is $0.005m$. The coordinate origin is the center of the cutout. The mass density $\rho$ is $3300kg/m^3$; Young's modulus $E$ is $45GPa$ and Poisson's ratio $\nu$ is 0.20. A constant tensile stress $q = 45MPa$, parallel with $x$ axis, is applied on the left and right sides of the plate.

For quasi-static simulations, Kilic and Madenci [23] introduced an artificial damping in their Peridynamic model. The effectiveness of this method was then verified by Madenci and Oterkus [36]. Due to the similarity between SPH and PD [17], this artificial damping is introduced in this work to simulate the quasi-static loading condition.

To accurately analyze the stress concentration around the circular hole, we use a finer particle distribution with radius of $0.0005m$ within the dense zone ($r_d = 0.02m$) as shown in Figure 10. The radius of the particles in the coarse zone is $0.001m$. There are in total 13459 particles used in the numerical models, including 4604 fine particles. The support radius associated to each particle is set as 3 times the particle size.

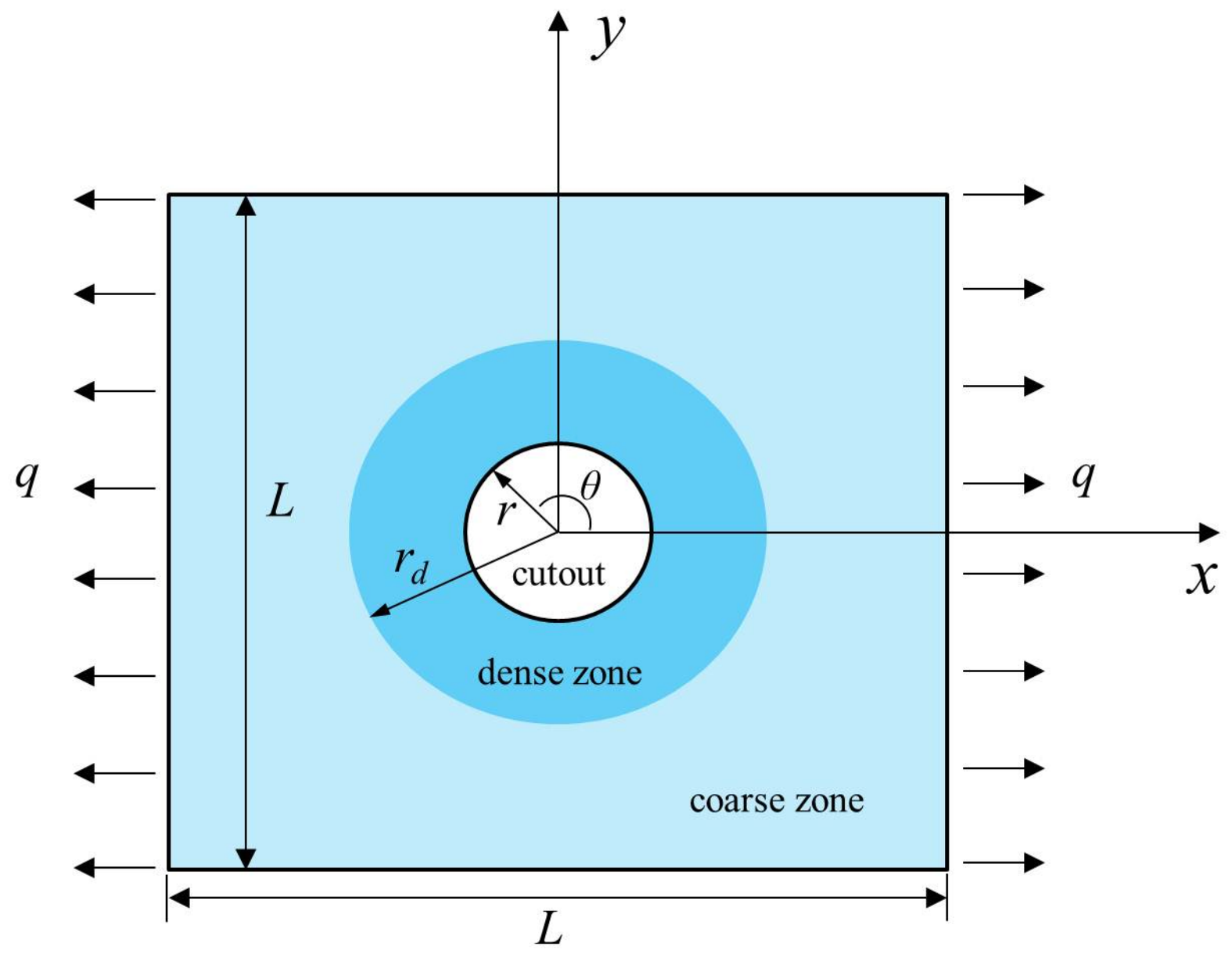

Figure 10: Geometry of an isotropic plate with a circular cutout under longitudinal tension

A material point located at ($-0.038m$, $-0.008m$) is monitored, and its displacement and stress variations with time are presented in Figure 11 and 12. From the figures, it can be easily concluded that the displacement and stress calculated by the DS-SPH model have relatively small fluctuation, and begin to stabilize much earlier than those from traditional SPH model.

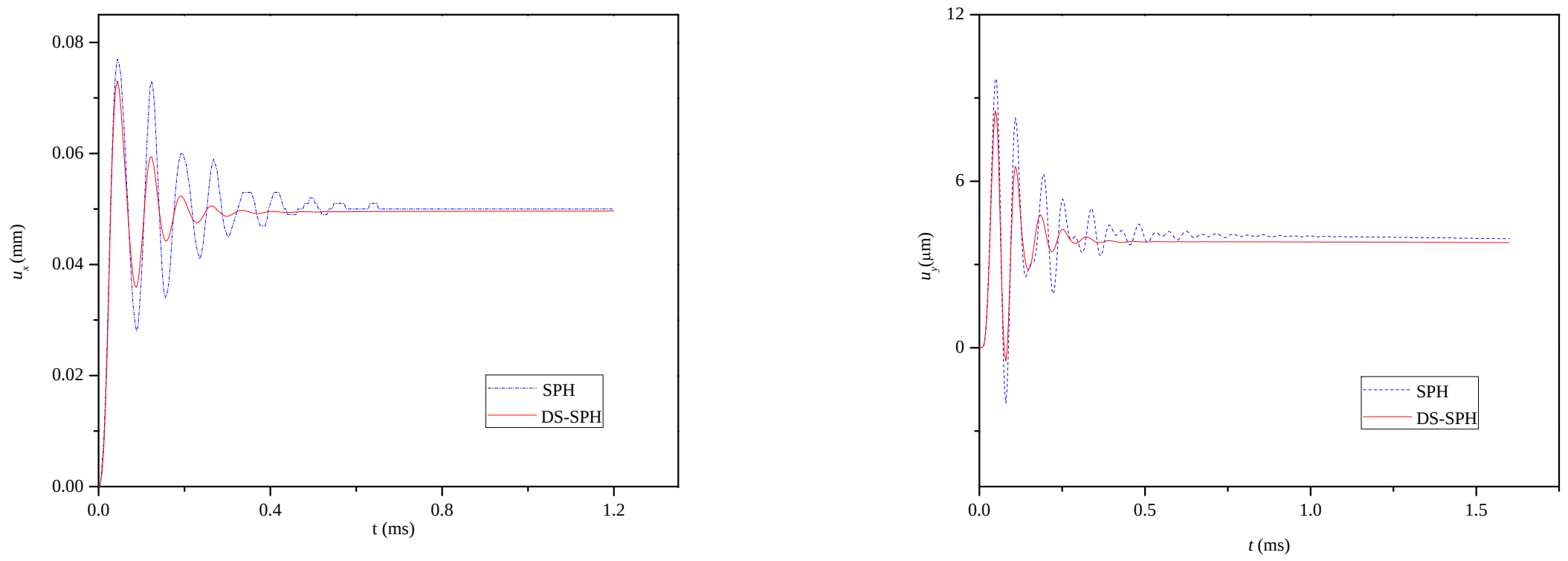

(a) $u_x$ (b) $u_y$

Figure 11: Time history of displacement

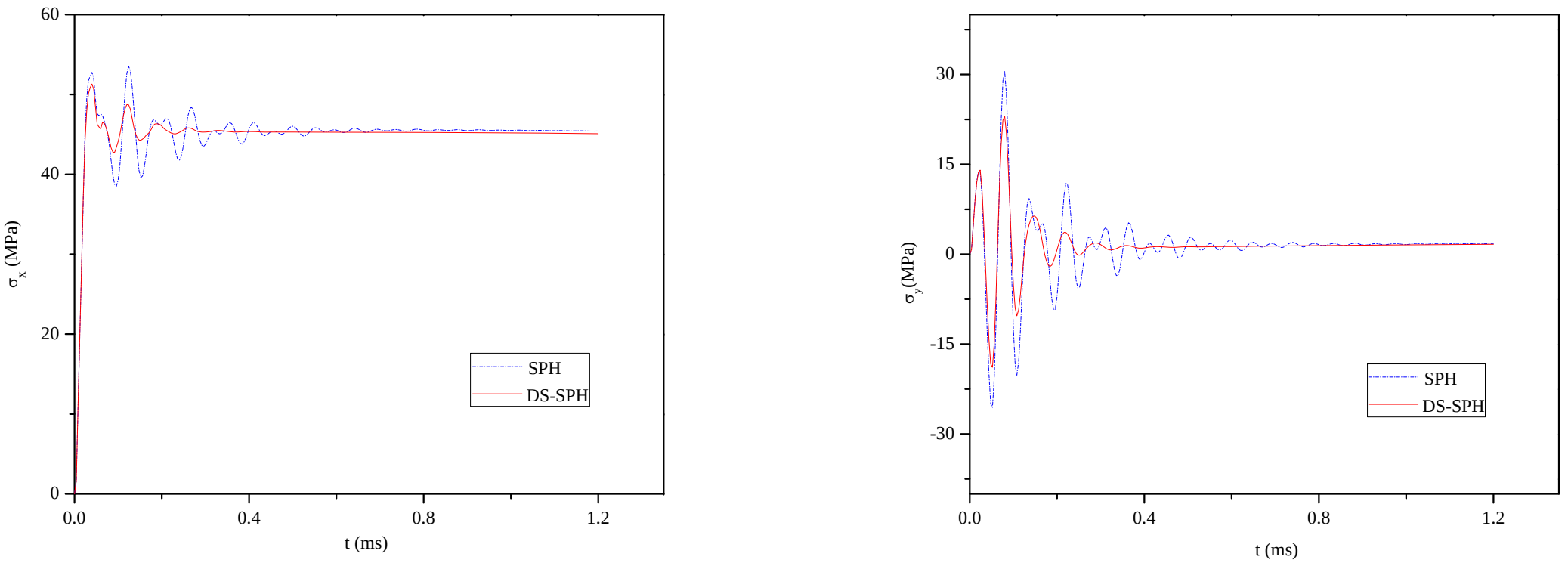

(a) $\sigma_x$ (b) $\sigma_y$

Figure 12: Time history of stress

Figure 13 shows the stress distribution of the isotropic plate with a circular hole under uniaxial tension at $t = 0.0016s$. In Figure 13 (a), due to the unbalanced forces, several red particles with relatively high stress can be found near the interface between the coarse and fine discretization. In relative terms, the stress distribution calculated by the DS-SPH model is smooth.

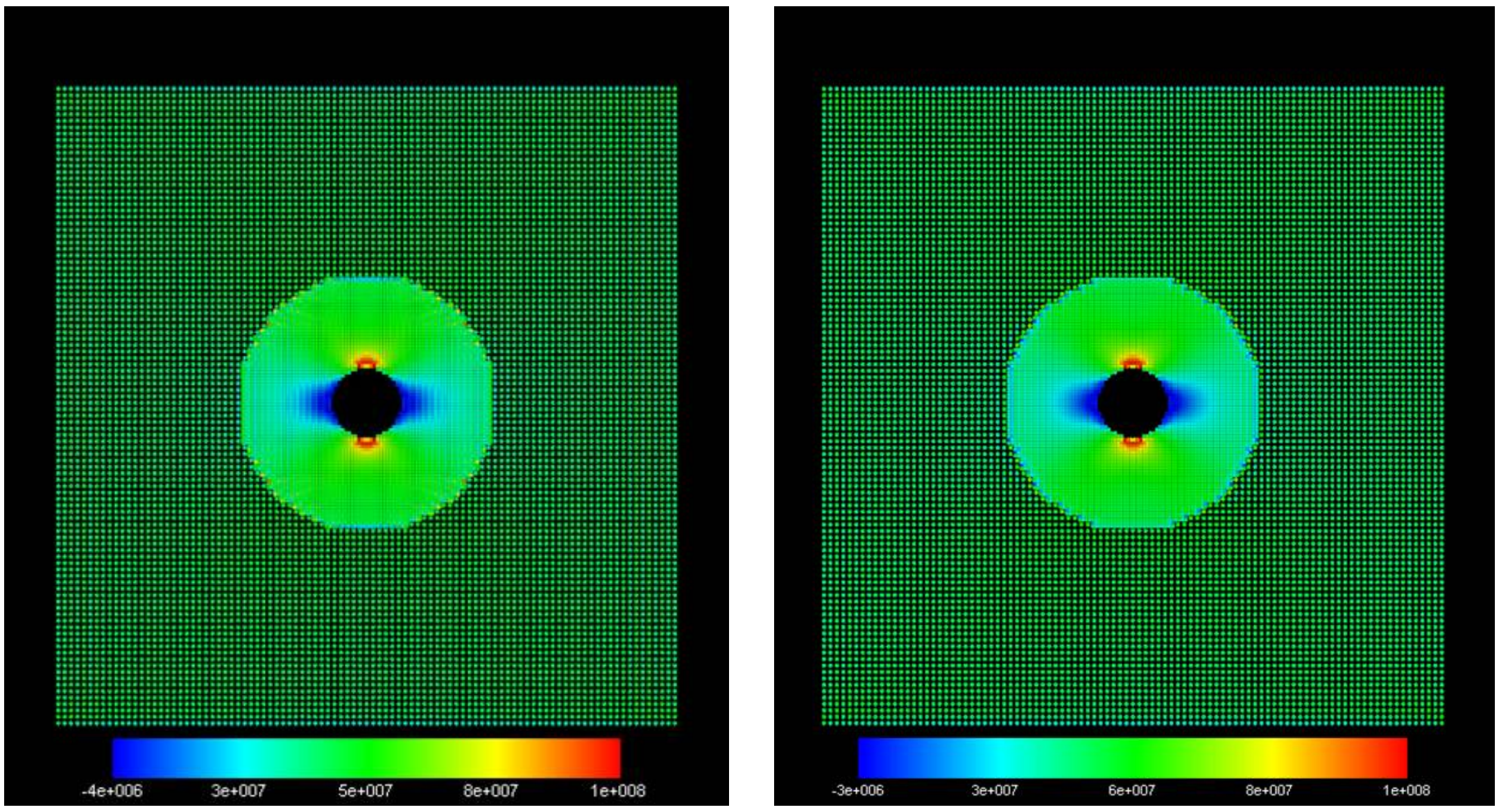

(a) SPH (b) DS-SPH

Figure 13: Stress contour of the isotropic plate

The analytic solutions of the stress field on the isotropic plate with a circular cutout can be obtained by the following equations [60]:

$$
\begin{aligned}
\sigma_x = & \frac{q}{2}[(1-\frac{a^2}{r^2})+(\frac{3a^4}{r^4}-\frac{4a^2}{r^2}+1)\cos(2\theta)]\cos^2\theta \\
& +\frac{q}{2}[(1+\frac{a^2}{r^2})-(\frac{3a^4}{r^4}+1)\cos(2\theta)]\sin^2\theta \\
& +\frac{q}{2}[(1-\frac{a^2}{r^2})(\frac{3a^2}{r^2}+1)\sin(2\theta)]\sin(2\theta)
\end{aligned} \tag{43}
$$

$$
\begin{aligned}
\sigma_y = & \frac{q}{2}[(1-\frac{a^2}{r^2})+(\frac{3a^4}{r^4}-\frac{4a^2}{r^2}+1)\cos(2\theta)]\sin^2\theta \\
& +\frac{q}{2}[(1+\frac{a^2}{r^2})-(\frac{3a^4}{r^4}+1)\cos(2\theta)]\cos^2\theta \\
& +\frac{q}{2}[(1-\frac{a^2}{r^2})(\frac{3a^2}{r^2}+1)\sin(2\theta)]\sin(2\theta)
\end{aligned} \tag{44}
$$

where $q$ is the uniform distribution load in horizontal direction, $a$ is the radius of the circular cutout, $r$ is the distance between the concerning point and the origin, $\theta$ is the angle relative to the positive $x$ axis. In this section, the stress distribution along the $x$ axis and $y$ axis are picked up to compare the numerical results from both SPH and DS-SPH models with analytic solutions.

Along the $x$ axis, $\theta = 0$, then the equations (43) and (44) can be simplified as:

$$\sigma_x = \frac{q}{2}(2+\frac{3a^4}{r^4}-\frac{5a^2}{r^2}) \tag{45}$$

$$\sigma_y = \frac{q}{2}(\frac{a^2}{r^2}-\frac{3a^4}{r^4}) \tag{46}$$

Along the $y$ axis, $\theta = \frac{\pi}{2}$, then the equations (43) and (44) can be simplified as:

$$\sigma_x = q(1 + \frac{a^2}{2r^2} + \frac{3a^4}{2r^4}) \tag{47}$$

$$\sigma_y = \frac{3q}{2}(\frac{a^2}{r^2} - \frac{a^4}{r^4}) \tag{48}$$

The analytic solution and numerical results are compared in Figure 14. The black solid line represented analytic solution, the blue triangles are the SPH results and the red rhombuses are the DS-SPH results. It is obvious that, the DS-SPH results are closer to the analytical solution than SPH results. Near the interface between the coarse and fine discretization ($x = 0.2m$), the stress calculated using the traditional SPH model has a considerable oscillation. Therefore, the traditional SPH method is not suitable for the problems with nonuniform particle distribution. Though a small oscillation exists, the DS-SPH results are acceptable.

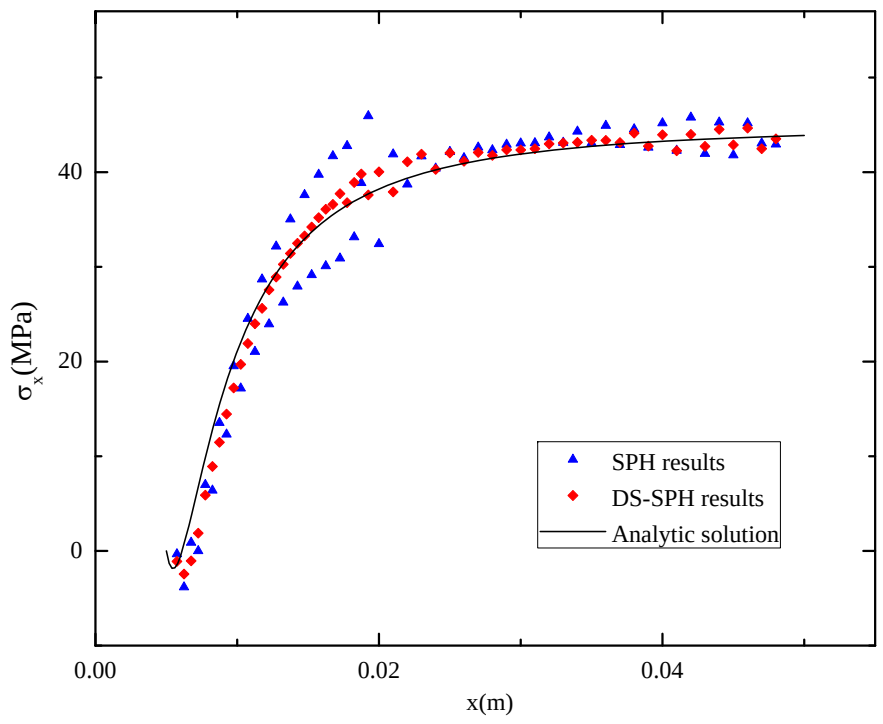

(a) $\sigma_x$ along $x$ axis

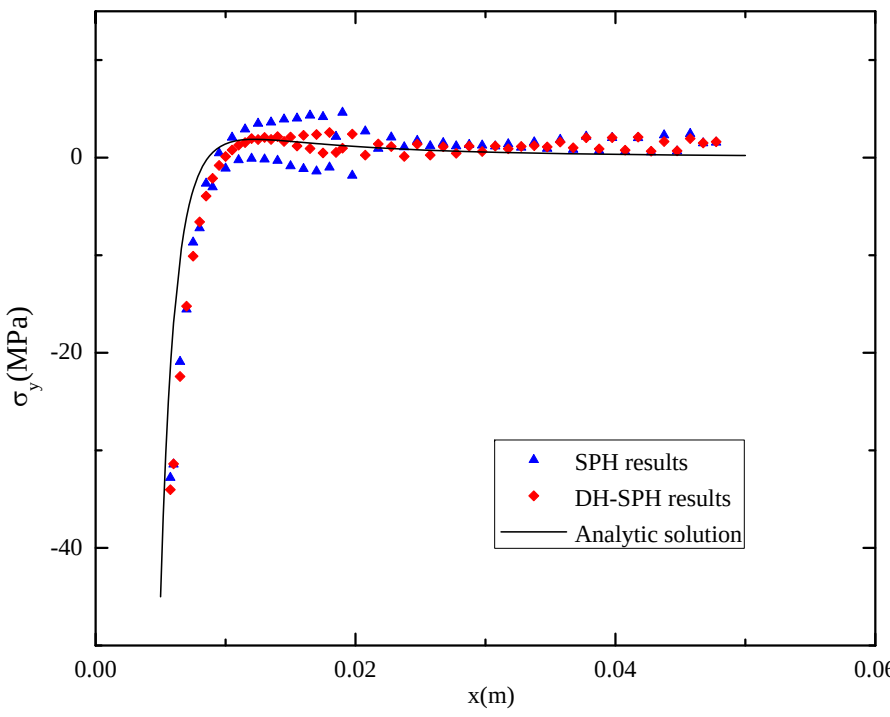

(b) $\sigma_y$ along $x$ axis

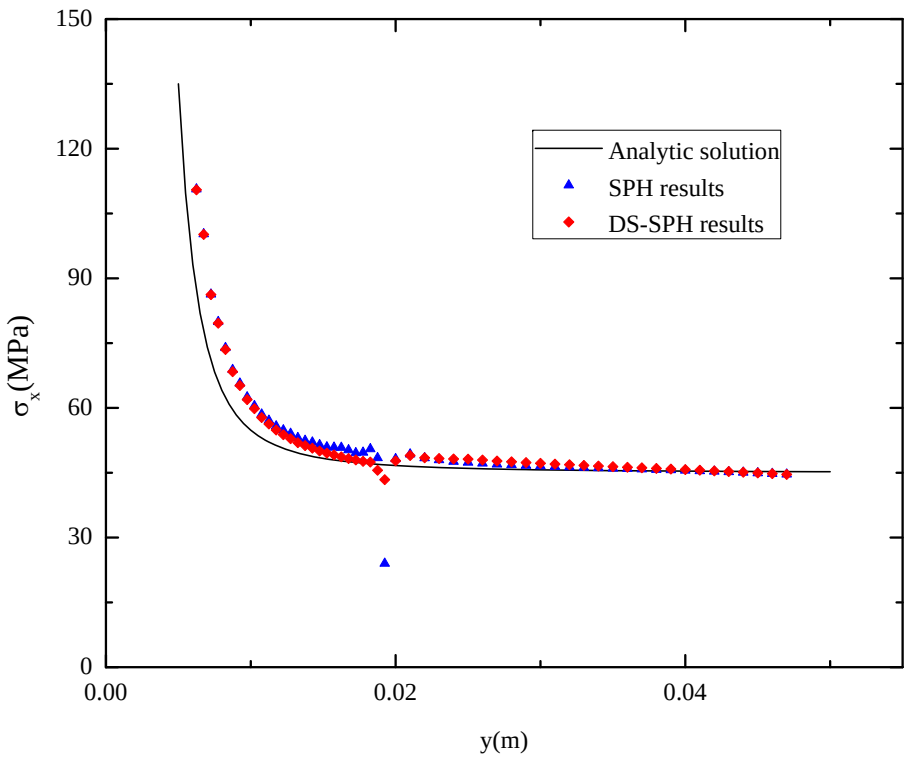

(c) $\sigma_x$ along $y$ axis

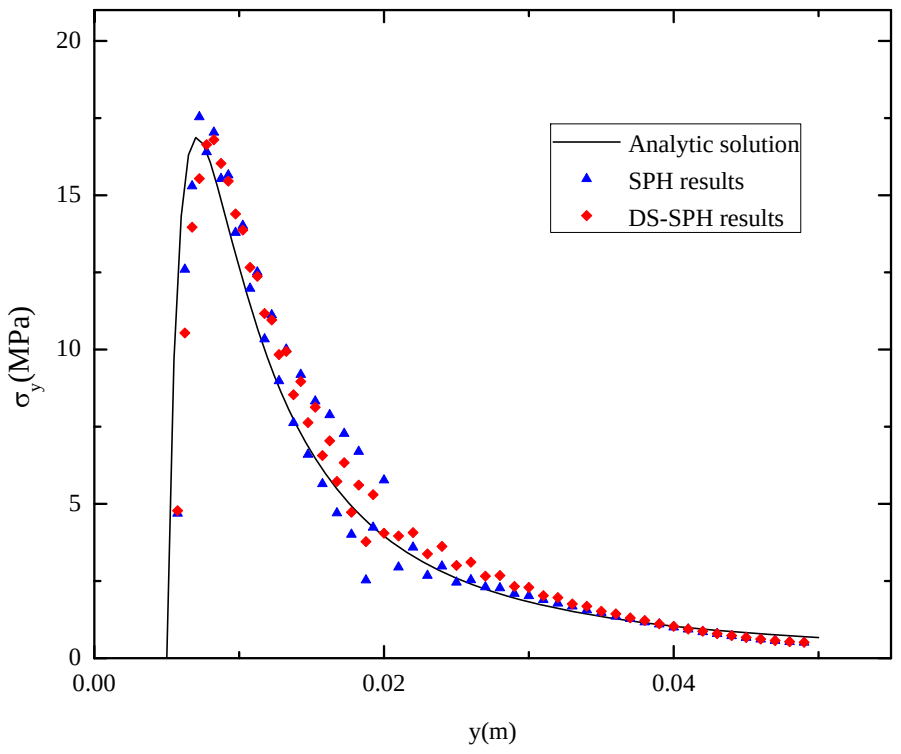

(d) $\sigma_y$ along $y$ axis

Figure 14: Stress distribution along the $x$ and $y$ axis of the isotropic plate

To compare the convergence rate of the DS-SPH and standard SPH method, the isotropic plate was discretized into 3509, 6165 and 13459 particles respectively. Figure 15 shows the stress distribution along $x$ axis simulated using the two models. The results obtained show good agreement with the analytical solution in particular for the fine distribution with 13459 particles.

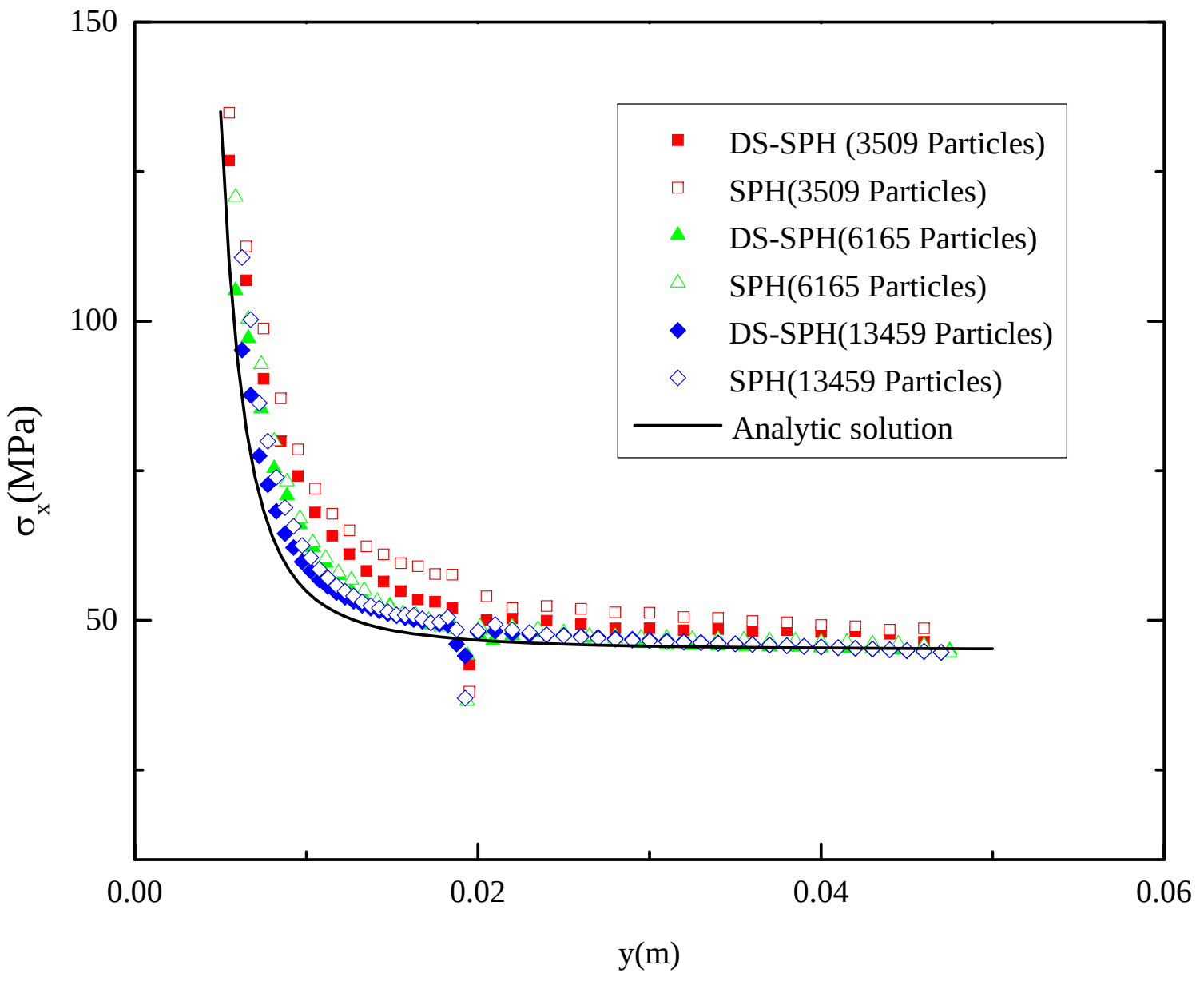

Figure 15: Normal stress $\sigma_x$ along $x = 0$

Figure 16 shows the convergence curves of DS-SPH and SPH models at $t = 1.6ms$. It indicates that the error in the energy norm of the DS-SPH model is smaller than the standard SPH model. The convergence rate of the DS-SPH model is 0.47, a little larger than the convergence rate of the standard SPH model 0.40.

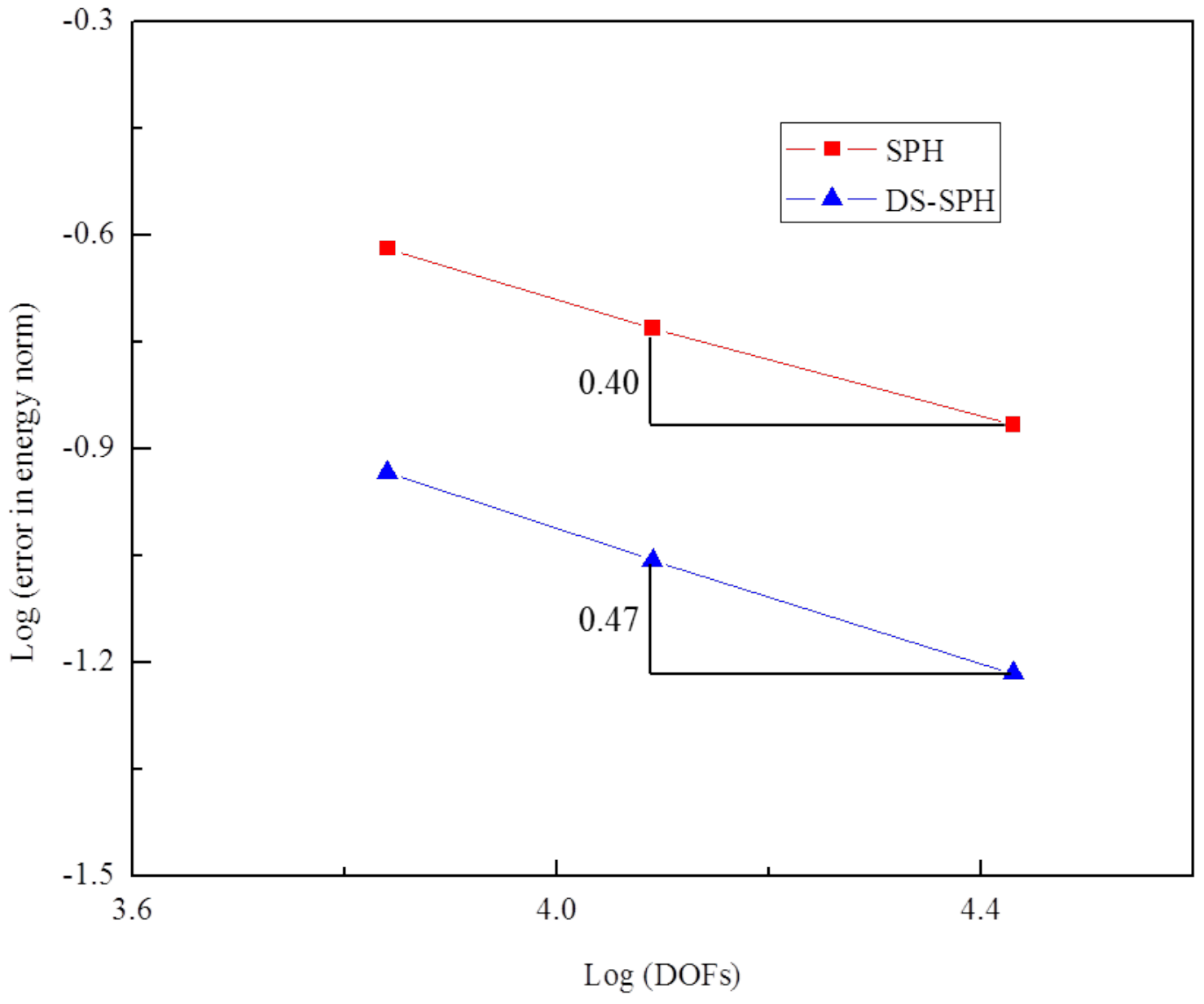

Figure 16: Convergence results in error norm

## 6.4 Block of material with a spherical cavity under radial extension

In this section, the DS-SPH model is applied to simulate a 3D problem. As shown in Figure 17, the length, width and height of the block are all $1m$, and the radius of the spherical cavity is $0.15m$. The Young's modulus, Poisson's ratio and mass density of the material are respectively $200GPa$, 0.25 and 7850 $kg/m^3$. In this problem, a spherical cavity inside a large block of material is subjected to radial extension. The block is free of loading on its outer surfaces. The radial displacement on the surface of the cavity is $0.001m$.

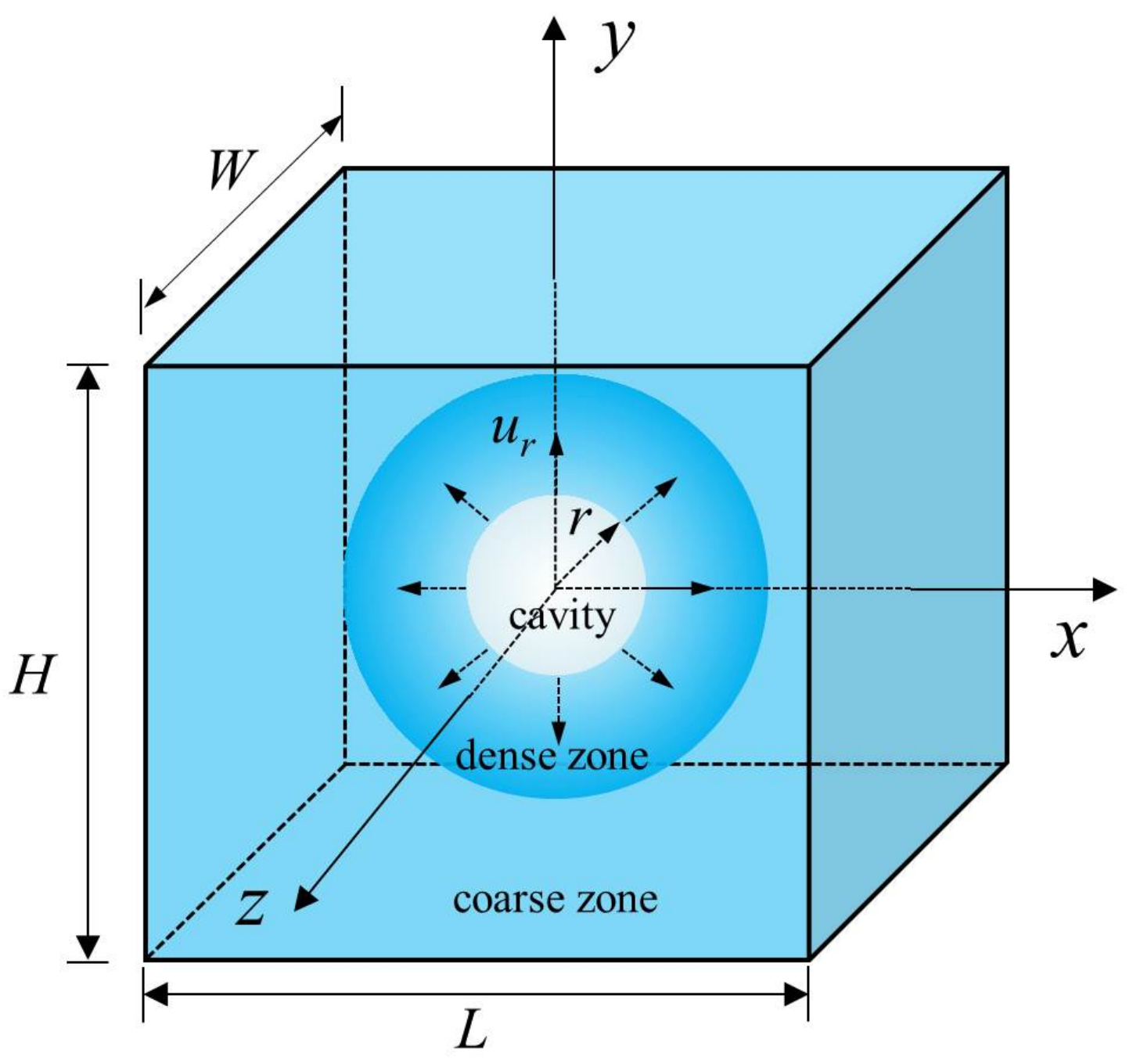

Figure 17: Geometry of a block of material with a spherical cavity under radial extension

A fine discretization is used in the zone around the cavity (the radius of this zone is 0.25m and the particle distance is 0.025m) while a coarse discretization is used elsewhere (the particle distance is 0.05m). The support size is set to 3 times the particle distance.

A material point located at $x_0$ $(0.45m, 0.35m, 0.3m)$ is monitored; the time histories of its displacement components $u_x$, $u_y$, $u_z$ are shown in the Figure 18. Under the radial extension, the displacement of the collocation point increases fast at the beginning, then slightly decreases, and eventually converges.

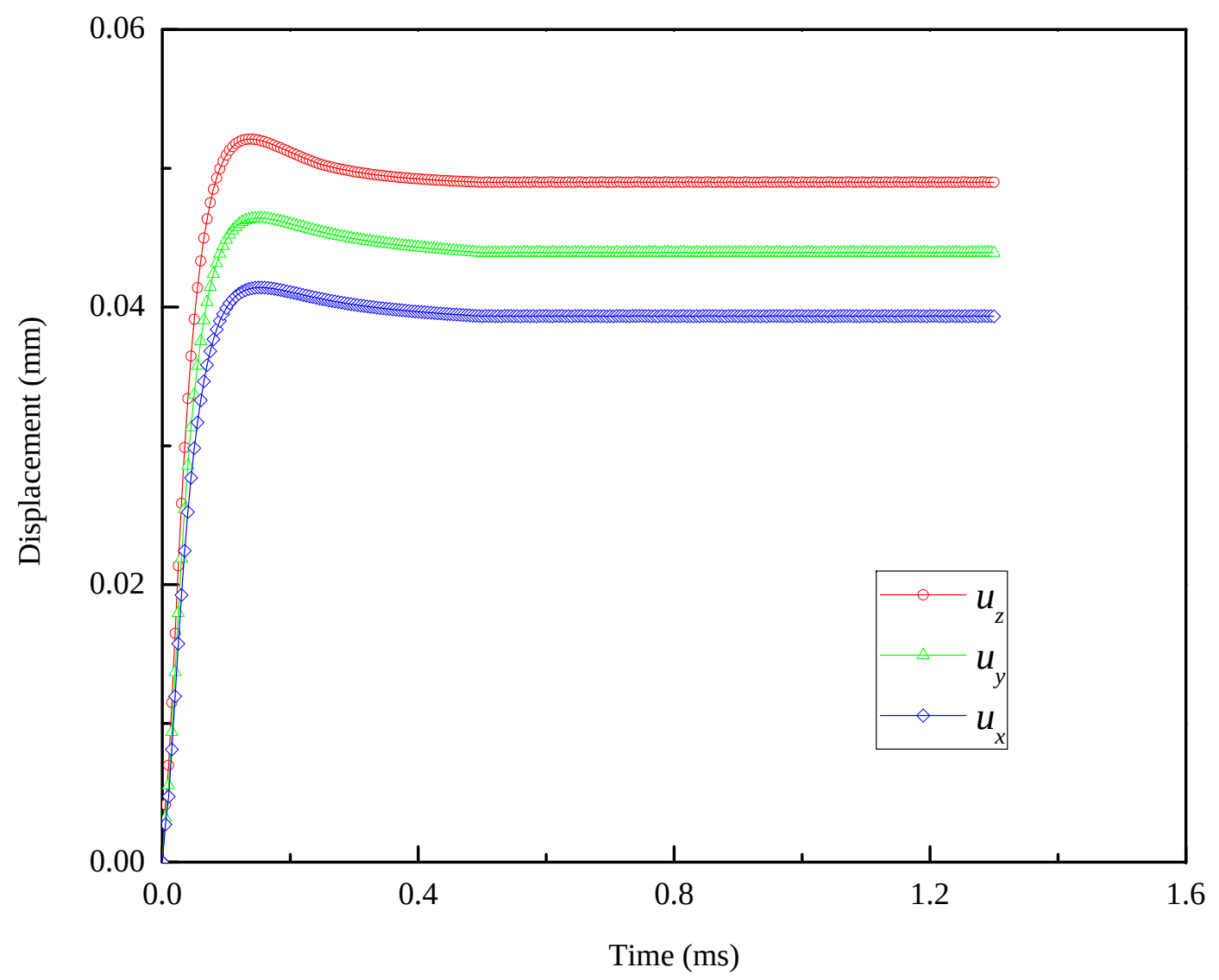

(a) Displacement

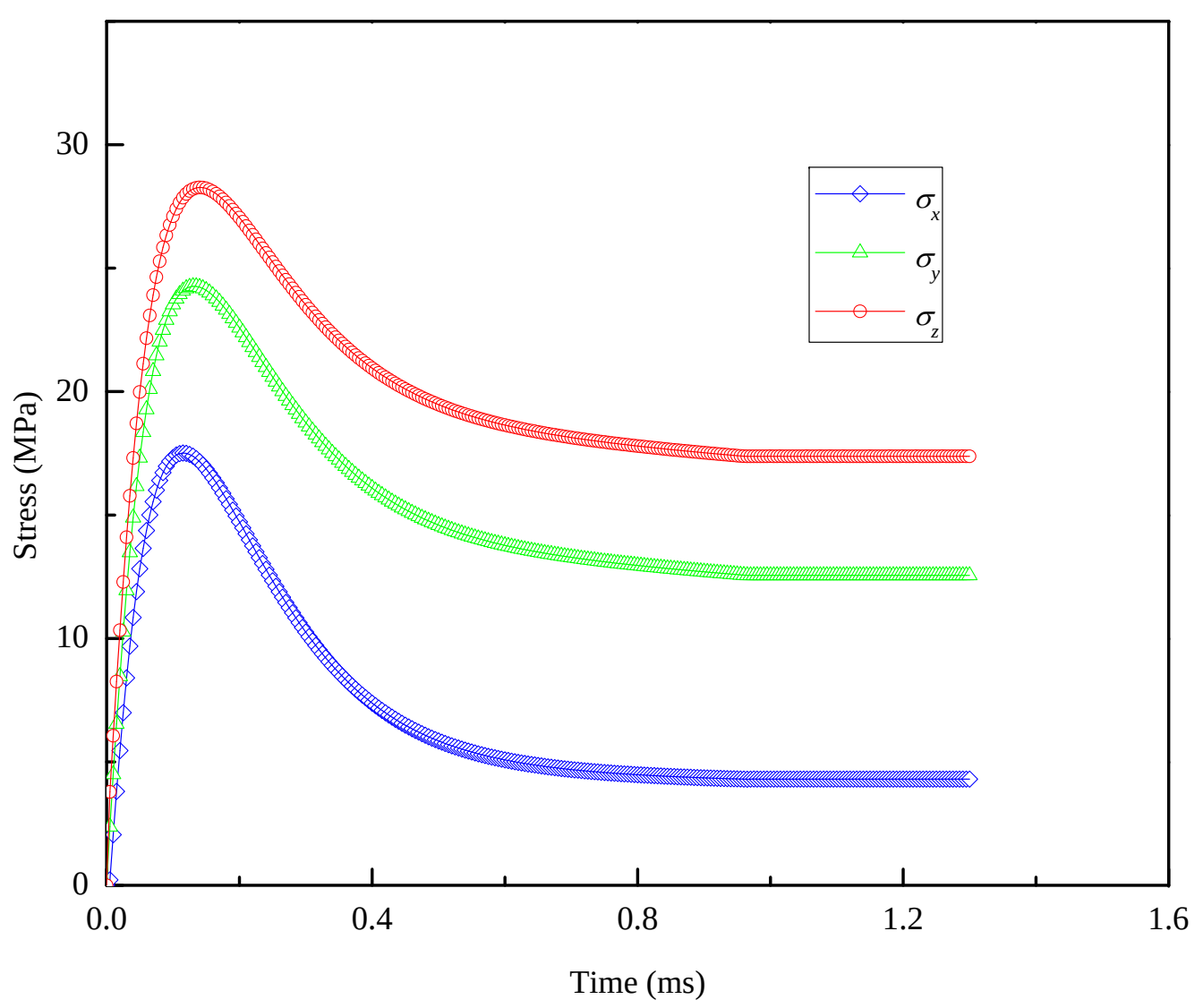

(b) Stress

Figure 18: Convergence of displacement stress components of the collocation point at $x_0$

After reaching the steady-state condition, the displacement and stress at each material point are shown in Figure 19. 1/4 of the block is cut off in order to show the displacement and stress distribution inside the block.

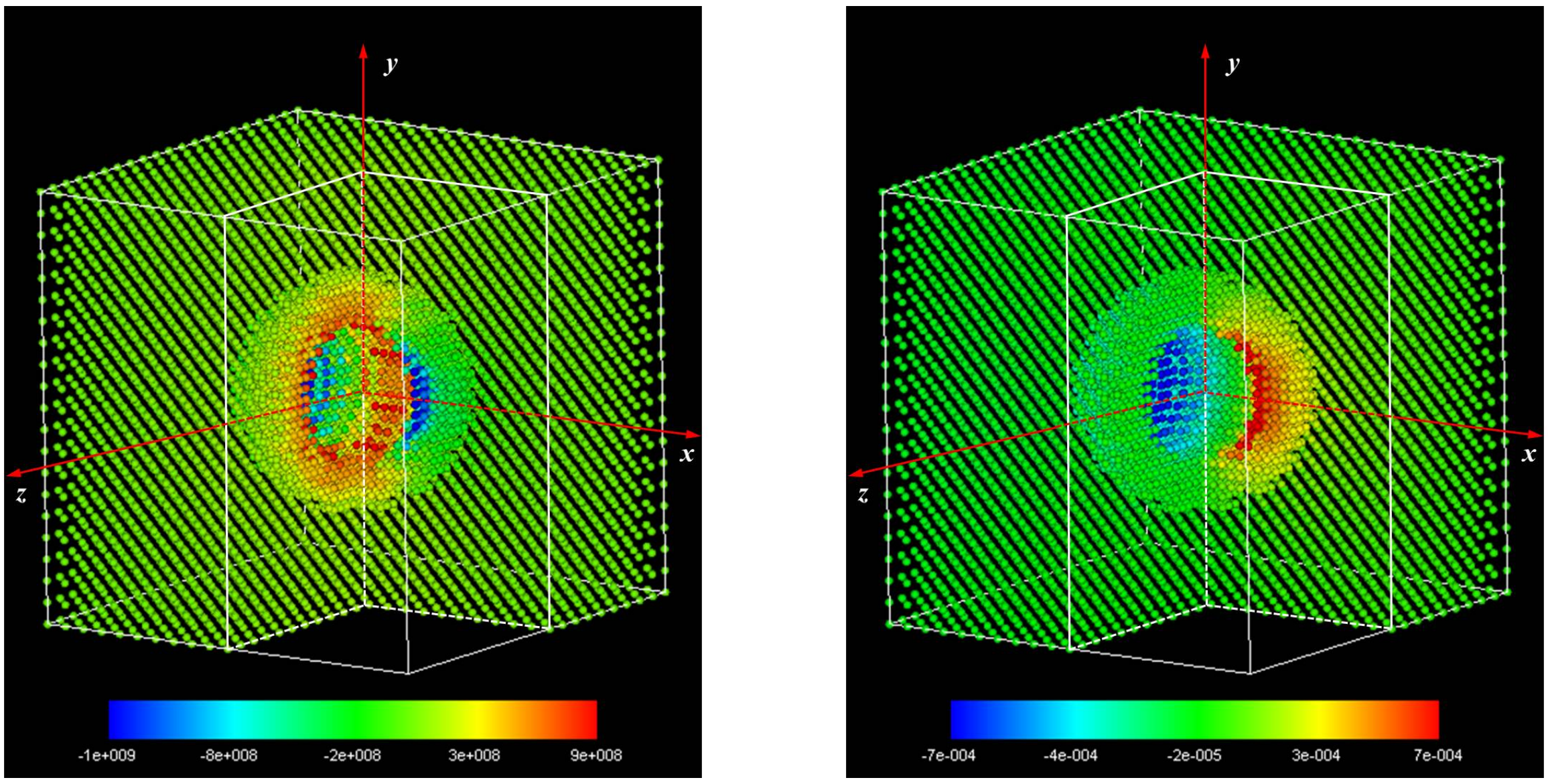

(a) $u_x$ (b) $\sigma_x$

Figure 19: Displacement and stress nephogram of the elastic block when $t = 1us$

To quantitatively analyze the results, the radial displacements and stress in the $x$-direction are compared with the analytical solution given by equation (49) and (50).

$$u_r = \frac{a^2}{r^2} u^* \tag{49}$$

$$\sigma_x = \frac{E}{(1+\nu)(1-2\nu)}[(1-\nu)\frac{\partial u_x}{\partial x} + \nu\frac{\partial u_y}{\partial y} + \nu\frac{\partial u_z}{\partial z}] \tag{50}$$

where $a$ is the radius of the cavity, $r$ is the distance between the concerning particle to the origin, $u^*$ is the displacement boundary applied on the surface of cavity, $E$ is the Young's modulus and $\nu$ is the Poisson's ratio.

Figure 20 shows the $u_x$ and $\sigma_x$ distribution along the $x$ axis. The results of the DS-SPH and standard SPH simulation with different number of particles are compared with the analytic solutions. It is obviously that the simulated results match the analytic solution in particular for the fine distribution with 24296 particles. Figure 21 shows the convergence curves of DS-SPH and SPH models at $t = 1.3ms$. It indicates that with the same particle distribution, the error in the energy norm of the DS-SPH model is smaller and its convergence rate is larger compared with SPH results.

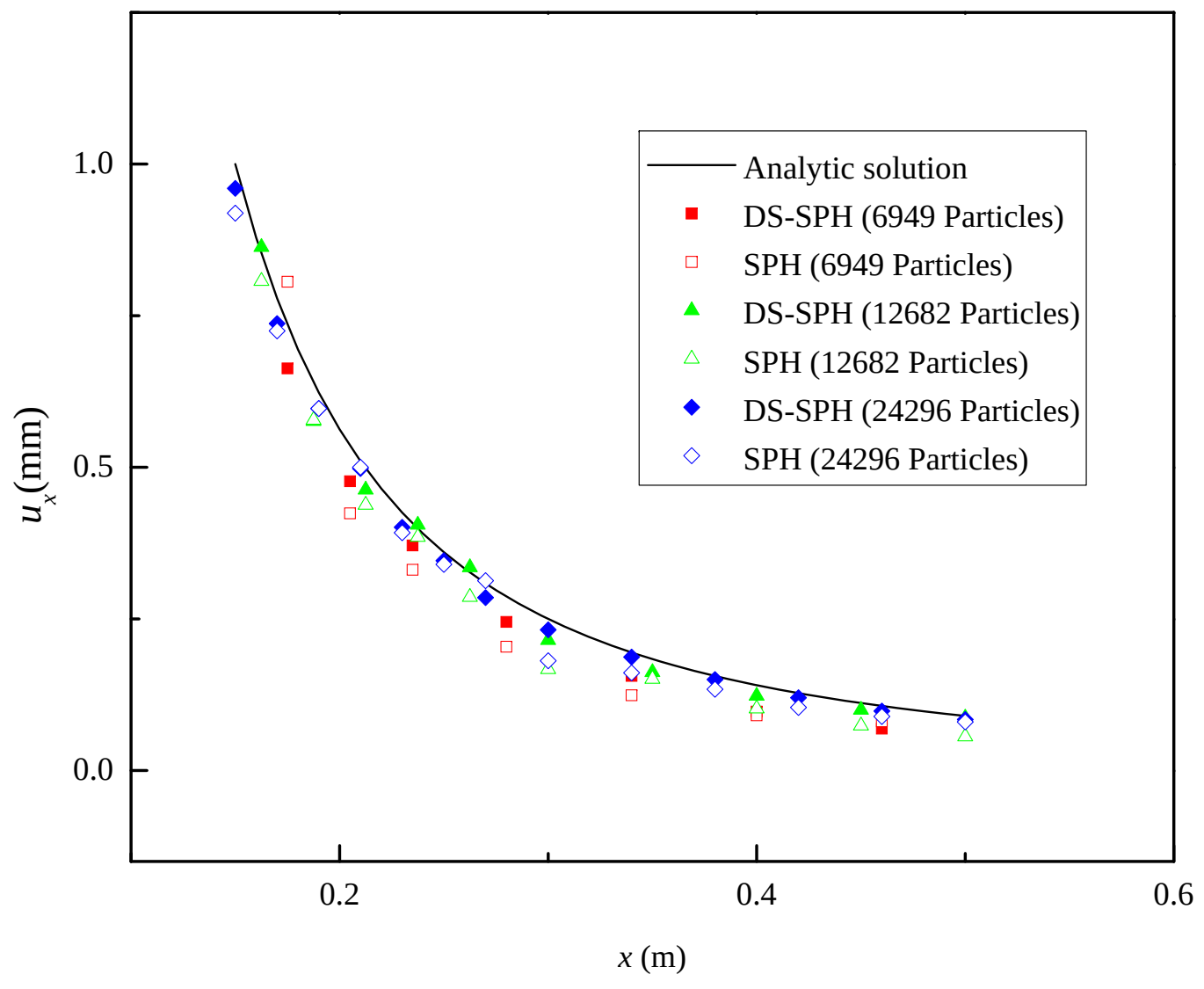

(a) $u_x$

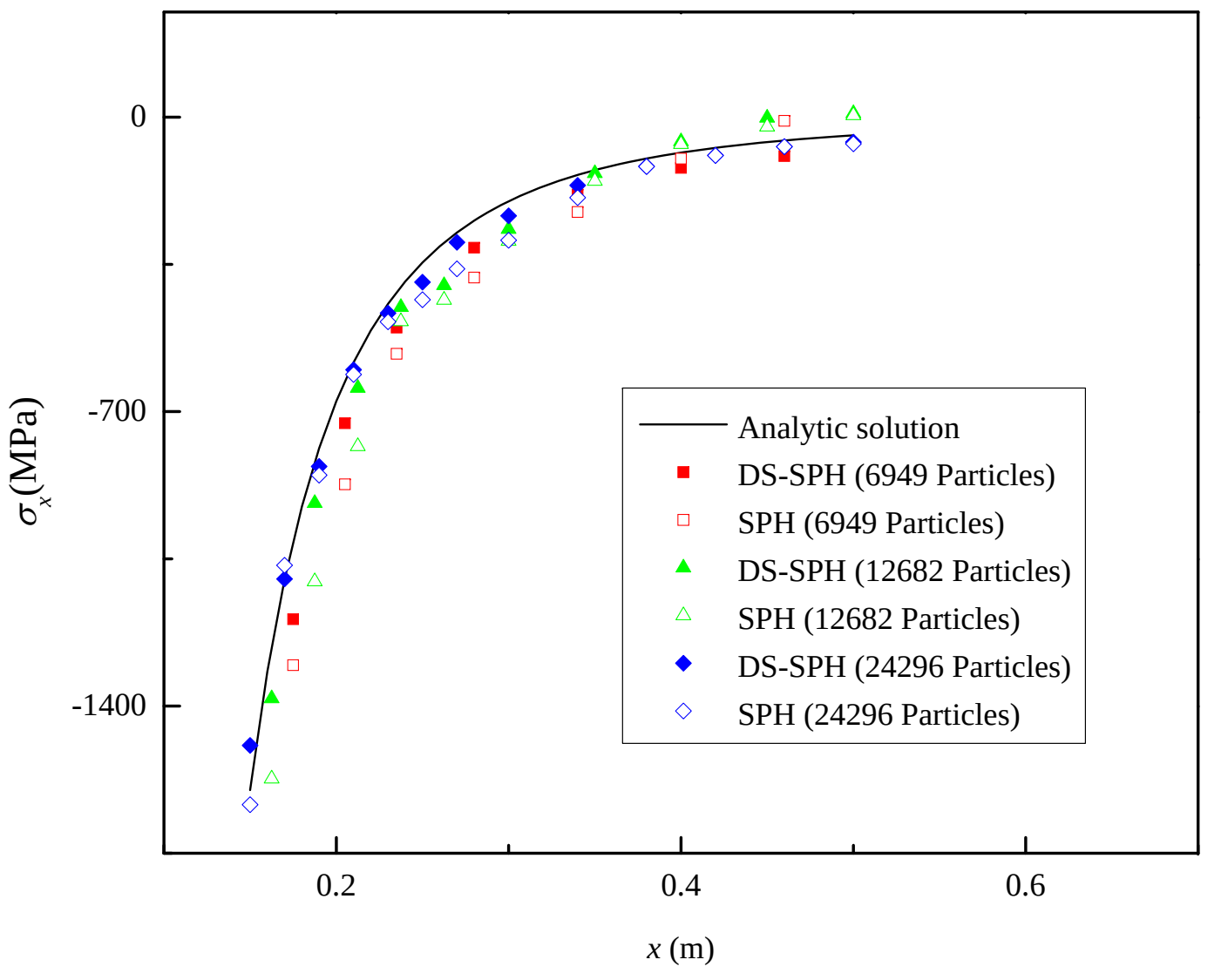

(b) $\sigma_x$

Figure 20: $u_x$ and $\sigma_x$ variation along $x-$axis

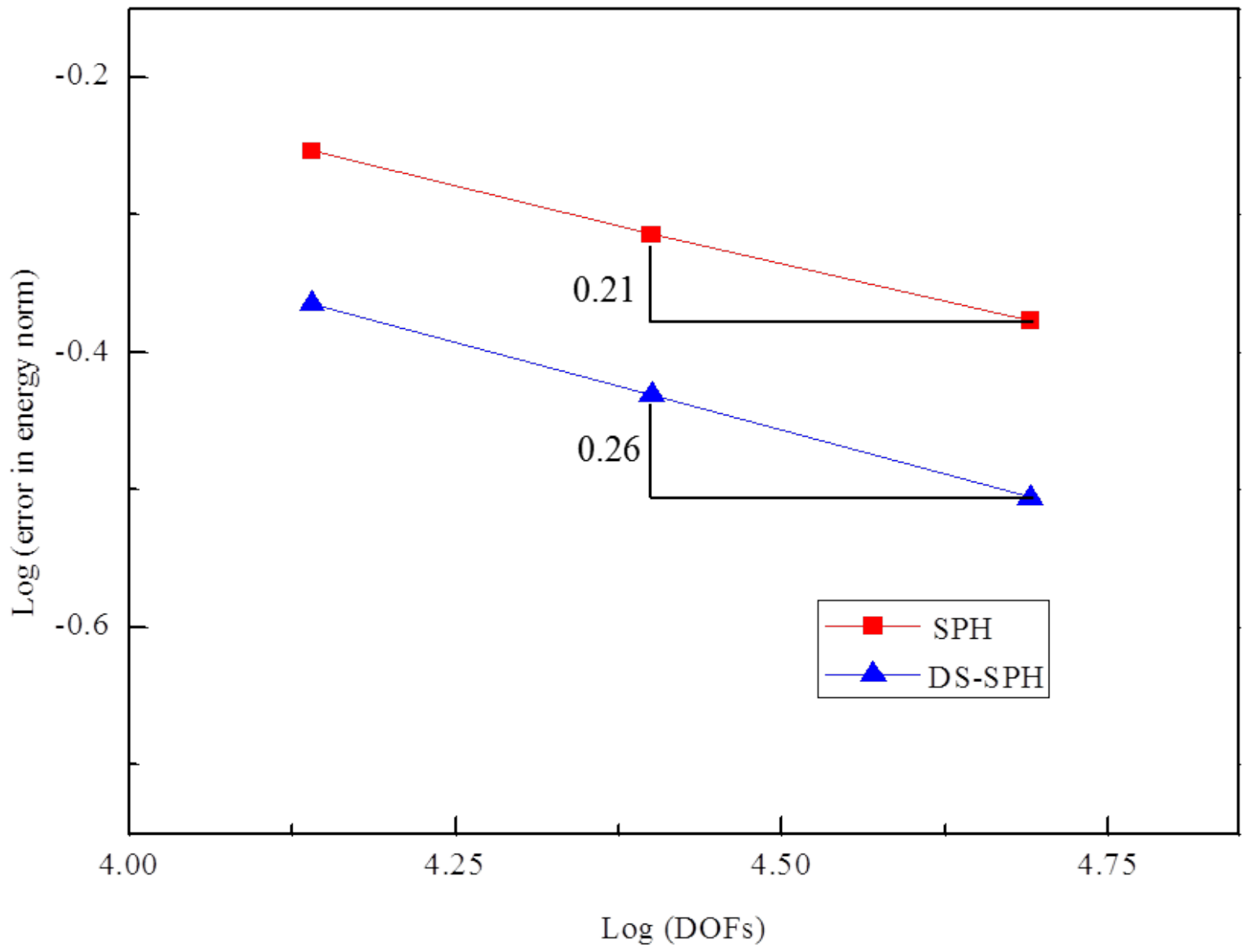

Figure 21: Comparison of convergence of DS-SPH with SPH

The case studies presented above shows that the key advantage of the proposed DS-SPH method is its ability to deal with non-uniform discretizations which are very important for computational efficiency. Its computational accuracy and convergence rate are better than the standard SPH solution. The significant feature of the meshfree particle methods is the ability to deal with the complex problems such as fracture modeling and multiscale analysis [33, 46, 47, 59]. Therefore, the next step of this work is to extend this approach to model fracture in the multiscale framework.

# 7 Conclusion

This work presents a general method to deal with the issue of varying supports and unbalanced internal force. A new SPH method with Lagrangian kernel and dual-support is proposed, and the derivation of the governing equations in traditional SPH method is modified. Based on the new SPH method, a numerical solution is formulated. Several numerical examples are simulated which verify the accuracy of the modified SPH model, and highlight its advantages over the traditional SPH model. The presented model also shows the potential in multiscale analysis where the models at different length scales can be bridged by distributing particles with different sizes and variable supports.

# 8 Acknowledgement

The authors acknowledge the supports from ERC-CoG(Computational Modeling and Design of Lithiumion Batteries (COMBAT)).